\documentclass[ twocolumn]{aastex701}

\usepackage{array}
\usepackage{longtable}
\usepackage{longtable}
\usepackage{amsmath}
\begin{document}

\title{A FUV–optical approach for studying hierarchical star formation in nearby galaxies with UVIT}

\shorttitle{UVIT FUV-optical study of hierarchical star formation }
\shortauthors{Ananthu et al.}

\author[0009-0002-9958-9494]{Sanal Ananthu}
\email{ananthusanal56@gmail.com}
\affiliation{Indian Institute of Astrophysics, II Block, Koramangala, Bengaluru 560034, India}

\author[0009-0008-1250-6128]{Gairola Shashank}
\email{shashank.gairola@iiap.res.in}
\affiliation{Indian Institute of Astrophysics, II Block, Koramangala, Bengaluru 560034, India}
\affiliation{Pondicherry University, R.V. Nagar, Kalapet, 605014, Puducherry, India}

\author[0000-0002-5331-6098]{Smitha Subramanian}
\email{smitha.subramanian@iiap.res.in}
\affiliation{Indian Institute of Astrophysics, II Block, Koramangala, Bengaluru 560034, India}
\affiliation{ Leibniz-Institut fur Astrophysik Potsdam (AIP), An der Sternwarte 16, D-14482 Potsdam, Germany}

\author[0009-0009-0141-6650]{Rao C. Jayanth}
\email{jayanthraoc@gmail.com}
\affiliation{Indian Institute of Astrophysics, II Block, Koramangala, Bengaluru 560034, India}

\author[0000-0001-5944-291X]{Shyam H. Menon}
\email{smenon@flatironinstitute.org}
\affiliation{Center for Computational Astrophysics, Flatiron Institute, 162 5th Avenue, New York, NY 10010, USA}
\affiliation{Department of Physics and Astronomy, Rutgers University, 136 Frelinghuysen Road, Piscataway, NJ 08854, USA}

\author[0000-0003-4531-0945]{Chayan Mondal}
\email{mondalchayan1991@gmail.com}
\affiliation{S. N. Bose National Centre for Basic Sciences
Block-JD, Sector-III, Salt Lake, Kolkata-700106, India}
\affiliation{Academia Sinica Institute of Astronomy and Astrophysics (ASIAA), No. 1, Section 4, Roosevelt Road, Taipei 10617, Taiwan}

\author[0009-0008-1606-497X]{Sreedevi Muraleedharan}
\email{sreedevim1909@gmail.com}
\affiliation{Indian Institute of Astrophysics, II Block, Koramangala, Bengaluru 560034, India}
\affiliation{Department of Physics, Indian Institute of Science Education and Research, Tirupati, Yerpedu, Tirupati - 517619, Andhra Pradesh, India}

\correspondingauthor{Sanal Ananthu}
\email{ananthusanal56@gmail.com}

\begin{abstract}
Young star-forming clumps (SFCs) emit strongly in the ultraviolet (UV), making UV imaging ideal for detecting them. The Ultraviolet Imaging Telescope (UVIT) onboard AstroSat, with 1.5\arcsec~resolution, has enabled the characterization of recently formed (up to 300 Myr) SFCs on tens of parsec scales in nearby galaxies. The spatial distribution of SFCs with different ages can provide insights into the hierarchy of star formation. This study presents a semi-novel approach to characterize SFCs in two nearby spiral galaxies, NGC 5457 and NGC 1313, by combining UVIT FUV data with g-band data from the Dark Energy Camera Legacy Survey (DECaLS). We tested and optimized our method on NGC 5457 and showed that after proper background subtraction, the FUV$-$g color of SFCs can serve as an equally reliable age indicator as the widely-used FUV$-$NUV color. Next, we parametrized the star formation hierarchy in NGC 5457 using the two-point correlation function (TPCF) and found good agreement between the hierarchy parameters derived using FUV$-$NUV and FUV$-$g based ages. Using our FUV$-$g based SFC ages, we also constrained the global hierarchy parameter of NGC 1313 for the first time. The development of our FUV$-$g based method is motivated by the fact that the NUV channel of the UVIT is not operational, and there is a wealth of archival UVIT FUV-only observations of nearby galaxies. This work demonstrates the potential of our method in constraining the SFC' ages and investigating hierarchical star formation in nearby galaxies using FUV and optical observations.

\end{abstract}
\section{Introduction}\label{sec1}
\begin{figure*}[htbp]
\centering
    \includegraphics[width=0.47\linewidth]{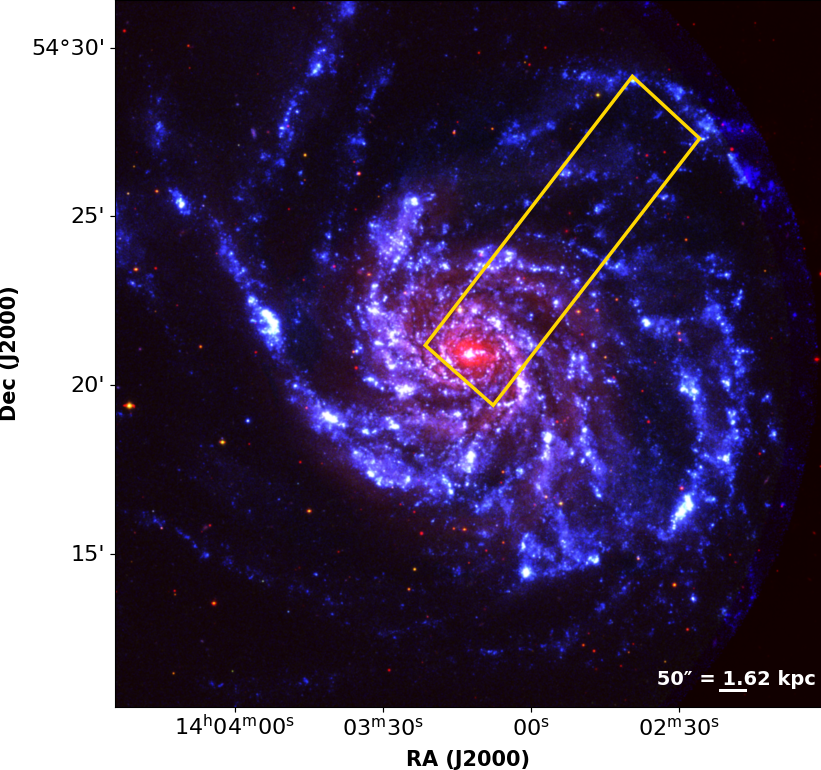}
    \hfill
    \includegraphics[width=0.48\linewidth]{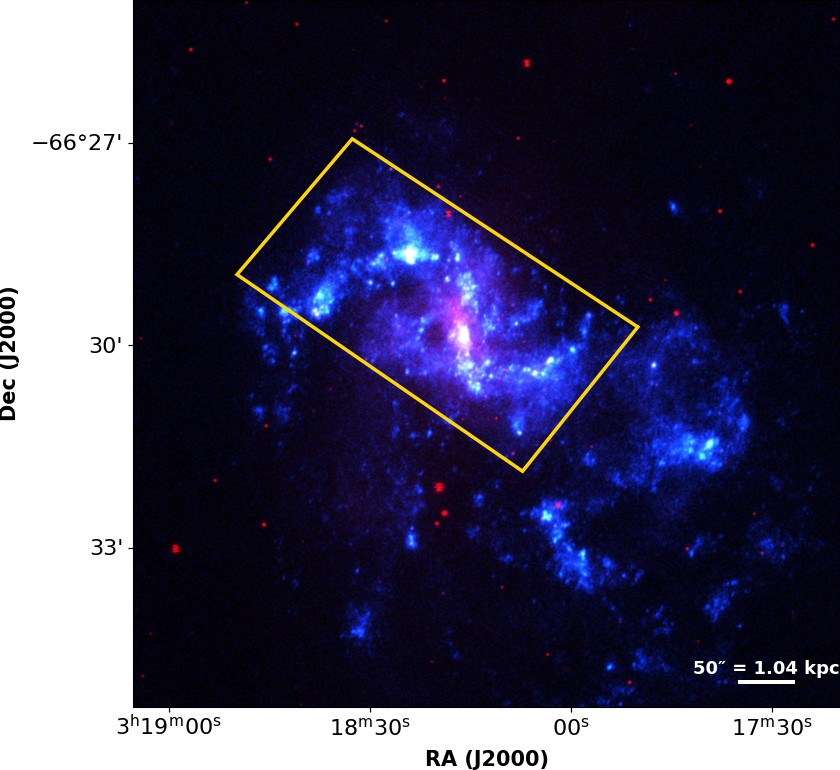}
\caption{Left: False-color composite image of NGC 5457, where blue, yellow, and red represent FUV, NUV, and g-band emissions, respectively. Right: False-color composite image of NGC 1313, where blue and red represent FUV and g-band emissions, respectively. In both panels, the golden box indicates the HST FoV ($3\arcmin \times 3\arcmin$ per pointing), adopted from \cite{menon2021dependence}. NGC 5457 is covered by four pointings ($36~\mathrm{arcmin}^2$), while NGC 1313 is covered by two pointings ($18~\mathrm{arcmin}^2$).}
\label{fig1}
\end{figure*}

Observations show that star-forming regions are arranged in self-similar, fractal-like structures where smaller and compact regions are iteratively nested within larger and more diffuse regions (\citealt{Elmegreen_1997,2001AJ....121.1507E,lada2003embedded,elmegreen2006hierarchical,gouliermis2017hierarchical}). The process of forming stars is predominantly governed by gravitational collapse and turbulent motions within molecular clouds which results in the reshaping of molecular clouds into hierarchical structures (\citealt{1981MNRAS.194..809L,1996ApJ...471..816E}). The star formation occurring in these clouds adopts the hierarchical structuring of the natal clouds. Studying the hierarchical star formation process is crucial for understanding how star-forming regions within galaxies and galaxies themselves as a whole, evolve in an interconnected, multi-scale way \citep{Elmegreen_2010}. It is well known that the hierarchical patterns of star formation are observable only in the young star-forming regions, as these regions are unaffected by the many factors that gradually disperse the hierarchies within galaxies, such as stellar feedback, stellar drift, galactic dynamics and tidal forces \citep{zhang2001multiwavelength,sanchez2010fractal,grasha2017hierarchicala,grasha2017hierarchicalb,menon2021dependence,shashank2025tracing}. This makes it essential to differentiate between young and old star-forming regions when investigating hierarchical star formation.

\indent Multi-wavelength observations from the Hubble Space Telescope (HST) with the Legacy ExtraGalactic Ultraviolet Survey (LEGUS) program has greatly enriched our current understanding of hierarchical star formation in external galaxies (\citealt{calzetti2015legacy,elmegreen2014hierarchical,grasha2017hierarchicala,grasha2017hierarchicalb,gouliermis2017hierarchical,menon2021dependence}). These studies have helped constrain the hierarchy parameters (a set of quantities derived by analyzing the hierarchical distribution of star-forming regions within a galaxy) for a number of nearby galaxies. These parameters include correlation length (the largest scale up to which the spatial distribution of star-forming regions traces the hierarchical (self-similar or fractal) structure of the ISM produced by turbulence-driven fragmentation), hierarchy dispersal timescale (the time taken for the dispersal of hierarchically distributed star-forming regions), and fractal dimension (which quantifies the complex, space-filling nature of any distribution). However, some of the galaxies studied in these papers have also been hampered by the limited spatial coverage of the HST. In this context, the Ultraviolet Imaging Telescope (UVIT), onboard AstroSat, offers a significant advantage due to its large 28\arcmin~field of view (FoV) which enables UV observations with full galaxy coverage for nearby galaxies. UVIT offers an angular resolution of 1.5\arcsec~and it has multiple broad and narrow band filters \citep{2006AdSpR..38.2989A,2012SPIE.8443E..1NK,2014SPIE.9144E..1SS}.

\begin{deluxetable*}{lcccccccc} \label{table1}
\tablewidth{\linewidth}
\tablecaption{The relevant properties and details of the UVIT observations of the galaxy sample \label{tab:uvit_summary}}
\tablecolumns{9}
\tablehead{
\colhead{Galaxy} & \colhead{$M_*$} & \colhead{D} & \colhead{i} & \colhead{P.A} & 
\colhead{FUV Exp} & \colhead{NUV Exp} & \colhead{$N_{\rm SFC}$ (FUV--NUV)} & \colhead{$N_{\rm SFC}$ (FUV--g)} \\
\colhead{}       & \colhead{($M_\odot$)} & \colhead{(Mpc)} & \colhead{(deg)} & \colhead{(deg)} &
\colhead{(s)} & \colhead{(s)}
}
\startdata
NGC 5457 & $1.9\times10^{10}$ & 6.7 & 18.0 & 39 & 3163(F148W) & 2987(N263M) & 1122 & 1208 \\ [3pt]
NGC 1313 & $2.6\times10^9$    & 4.3 & 51   & 14 & 4735(F148W) & NA  & NA & 1000 \\
\enddata
\tablecomments{The stellar mass ($M_*$), distance (D), inclination (i), and position angle (P.A) were taken from \citet{menon2021dependence}. FUV filter F148W corresponds to 1481 Å central wavelength and N263M corresponds to 2632 Å. NGC 1313 lacks NUV observations. The number of SFCs ($N_{\rm SFC}$) is based on a magnitude error cut of 0.1 mag for NGC 5457 and 0.15 mag for NGC 1313, across all the wavebands used.}
\end{deluxetable*}

\indent UV observations effectively trace recent star formation (\citealt{calzetti2013star,kennicutt1998global}) within $\sim$300 Myrs. The FUV$-$NUV color of young star-forming regions is an excellent proxy for their age, as it is sensitive to the presence of short-lived, massive O and B type stars. Many GALEX based studies \citep{bianchi2005recent, 2005ApJ...619L..79T, 2005ApJ...619L..67T, 2011ApJ...743..137B} and UVIT based studies \citep{Mondal_2021,10.1093/mnras/stac2285,shashank2025tracing} have used the FUV$-$NUV colors of the star-forming regions to derive their ages. Recently, \cite{shashank2025tracing} utilized the full galaxy coverage UVIT FUV (F148W filter), NUV (N242W or N263M filter) observations, and the FUV–NUV color-based age estimates to investigate hierarchical star formation in four nearby spiral galaxies - NGC 1566, NGC 5194, NGC 5457, and NGC 7793. Their study demonstrated the capabilities of UVIT in constraining the hierarchy parameters of nearby galaxies. By comparing their hierarchy parameters (derived using full galaxy coverage with the UVIT) against those derived in \cite{menon2021dependence} (derived using partial galaxy coverage with the HST), they also demonstrated that the global hierarchy parameters of a galaxy can be significantly different from the parameters derived with partial galaxy coverage. This suggests that, owing to the significant environmental variations that can exist within different regions of a galaxy, partial galaxy coverage makes it difficult to accurately derive its global hierarchy parameters. For NGC 1566 and NGC 5194, \cite{menon2021dependence} had already provided lower limits of correlation length using partial galaxy coverage HST data. However, these lower limits made it unclear whether there exists a correlation length for these galaxies, or the stellar hierarchy can span up to the scales comparable to the full galaxy size. \cite{shashank2025tracing} analyzed the spatial distribution of young ($<$ 10 Myr) star-forming clumps (SFCs) in these galaxies and provided exact values for the correlation lengths of NGC 1566 and NGC 5194. Their correlation length measurements were also consistent with the lower limits provided by \cite{menon2021dependence}. This also demonstrated that with full galaxy coverage, all spiral galaxies are expected to have their own characteristic correlation length, which is much smaller than the entire galaxy size. 

\indent There exists a wealth of archival UVIT observations of nearby galaxies, with full disk coverage, which provides an opportunity to derive reliable hierarchy parameters for a statistically large sample of nearby galaxies. It will help to understand the dependence of hierarchy parameters on host galaxy properties. However, the NUV channel of the UVIT has not been operational since 2018, and a lot of the nearby galaxies are limited to observations only in the UVIT FUV channel. This hinders the application of the methodology adopted in \cite{shashank2025tracing} for deriving the SFC ages using the FUV$-$NUV colour, and thereby the estimation of the hierarchy parameters of galaxies. To address this, we explore in this paper the feasibility of an alternative method using a combination of UVIT FUV and Legacy survey g-band observations to estimate the ages of the SFCs. This approach will enable us to investigate hierarchical star formation in those nearby galaxies for which only UVIT FUV observations are available.\ 

The aim of this paper is two-fold. First, we study the galaxy NGC 5457 in order to develop a semi-novel method for estimating the ages of the SFCs using FUV–g color. We validate our method by comparing the SFC ages and hierarchical properties of NGC 5457 derived using FUV–g color against those derived using FUV–NUV color. Second, we apply our newly established method to the galaxy NGC 1313, for which UVIT NUV data are not available, and only the lower limit of the correlation length has been derived in the literature (960 pc; \citealt{menon2021dependence}). We utilize the full galaxy coverage UVIT FUV observations of NGC 1313, identify young SFCs based on the FUV–g color-based ages, and derive the hierarchy parameters of this galaxy. In this work, we measured the exact correlation length of NGC 1313 for the first time.

The structure of the remaining paper is as follows: In Sections \ref{sec2} and \ref{sec3}, we describe the galaxy sample and data used in this study, respectively. Section \ref{sec4} covers the methods related to SFC identification, photometry, age estimation, and the investigation of star formation hierarchy in our galaxies. In Section \ref{sec5}, we discuss the results of our age estimation methods, investigate the age demographic of our sample galaxies, and derive their hierarchy parameters. Finally, we summarize our results and discuss future prospects in Section \ref{sec6}.

\section{Galaxy sample} \label{sec2}

The sample galaxies studied in this paper are NGC 5457 and NGC 1313. These two are nearby, spiral galaxies rich in young star-forming regions. \\

\indent NGC 5457, also known as the Pinwheel Galaxy, is located in the Ursa Major constellation. Morphologically classified as SAB(rs)cd, it is an extended UV (XUV) disk spiral galaxy with a well-defined ring structure and loosely wound spiral arms \citep{1991rc3..book.....D}. It contains many young star-forming regions spread across its spiral arms, making it an ideal candidate for investigating the spatial distribution of star formation and its hierarchical nature. NGC 5457 was previously studied for hierarchical star formation using HST (\citealt{menon2021dependence}). However, these HST data provided only partial spatial coverage of approximately 10\% of the galaxy. Using full coverage UVIT observations \cite{shashank2025tracing} explored the global view of hierarchical star formation in NGC 5457. Furthermore, \cite{yadav2021comparing} investigated the extended UV (XUV) disk nature of this galaxy, identifying significant differences between the physical properties of inner (galactocentric radius $\leq$ $R_{25}$) and outer ($\geq$ $R_{25}$) star-forming regions. Here, $R_{25}$ is the radius of the galaxy corresponding to B-band surface brightness of 25 mag arcsec$^{-2}$.

\indent NGC 1313, a barred spiral galaxy of type SBd, is located in the constellation Reticulum. It hosts numerous HII regions \citep{messa2021looking} and shows enhanced star formation activity in the southern spiral arm \citep{suzuki2013akari}. In \cite{menon2021dependence}, the hierarchy parameters of NGC 1313 could not be properly constrained due to its partial spatial coverage within the HST LEGUS survey, which covered only about 43\% of the galaxy. As a result, we only have a lower limit of correlation length (960 pc) for this galaxy. The availability of UVIT FUV data with complete coverage for this galaxy gives us an opportunity to apply our FUV$-$g color-based methodology for deriving its global hierarchy parameters. Some important properties of our sample galaxies are listed in Table \ref{table1}, and their color-composite images are shown in Figure \ref{fig1}.

\section{Data} \label{sec3}
 
\indent We used the archival UVIT observations of NGC 5457 and NGC 1313. UVIT is one of the payloads onboard AstroSat - India's first multi-wavelength space observatory \citep{2006AdSpR..38.2989A,2012SPIE.8443E..1NK,2014SPIE.9144E..1SS}. UVIT consists of two telescopes simultaneously capturing images in the FUV (1300–1800 \AA) and NUV (2000–3000 \AA) channels. Additionally, UVIT has a VIS (3200–5500 \AA) channel, which is primarily used for tracking purposes. It provides excellent spatial resolution ($<$1.5\arcsec) and a large field of view of approximately 28\arcmin~diameter.

\indent To obtain science-ready images of the two galaxies, we used CCDLAB - a UVIT data reduction pipeline \citep{postma2021uvit, 2017PASP..129k5002P}. Due to the movement of the sky scanning monitor (SSM) camera on AstroSat, and the motion of the satellite, during the course of the observations, a drift pattern is introduced in the images. The short-exposure VIS channel observations are used to compute the drift pattern in the UV images, and the pipeline's drift correction option is used to correct the drift. UVIT observations for a given filter are usually taken in different orbits and in different orientations with respect to the celestial coordinates. This introduces rotational and translational shifts in the images taken during different orbits. These shifts were corrected using CCDLAB’s user-interactive registration routine, which involves manually placing orbit-wise images on top of each other using a number of bright point sources in the field (usually, 3 or 4 such sources are sufficient). CCDLAB performs the registration process with sub-pixel accuracy, ensuring that any shifts between orbit-wise images stay under one pixel (i.e. 0.416\arcsec) \citep{postma2021uvit}. After registration, the aligned images were merged to produce a single master image. Next, point spread function (PSF) optimization was performed to correct any residual drift that had not been corrected in the drift correction option. Multiple bright sources in the observed field were selected, and a common PSF correction is computed and applied by CCDLAB. This ensures that a uniform PSF is observed across the 28\arcmin~FoV of the UVIT. Lastly, astrometry is performed by accessing the Gaia DR3 catalog to assign the World Coordinate System (WCS) to the final image.

\indent We utilized g-band (3900–5400 Å) images from the legacy imaging surveys. NGC 5457 was imaged as part of the Beijing-Arizona Sky Survey (BASS), and NGC 1313 was imaged as part of the Dark Energy Camera Legacy Survey (DECaLS). BASS employed the Bok 2.3m telescope at Kitt Peak, equipped with the 90Prime camera \citep{dey2019overview}. The camera features four \( 4k \times 4k \) CCDs, providing a FoV of 1.15 square degrees. BASS achieved a spatial resolution of $\sim$1.6\arcsec. DECaLS employed the 4m Blanco telescope at Cerro Tololo Inter-American Observatory. It is equipped with a 570 megapixel dark energy camera, providing a FoV of 3 square degrees. DECaLS achieved a spatial resolution of $\sim$1.3\arcsec. We obtained the co-added images of our galaxies from the Legacy Survey website\footnote{\url{https://www.legacysurvey.org/viewer}}. Photometric zero points for these images were derived by comparing fluxes with the Pan-STARRS1 point source catalog.\\

\indent The legacy survey was optimized for uniform sky coverage rather than targeting specific objects. Imaging large, extended galaxies, such as our sample galaxy NGC 5457, required mosaicking data from multiple adjacent exposures to achieve seamless coverage. This step ensured complete galaxy coverage while compensating for the inherent CCD gaps. We used the Montage software to perform the mosaicking \citep{jacob2009montage}. The FUV, NUV, and g-band images of NGC 5457 used in this work have angular resolutions of less than 1.4\arcsec, 1.1\arcsec, and 1.6\arcsec, respectively. For our analysis, the average PSF of all the images was matched and brought to a common resolution of 1.6\arcsec, i.e, the angular resolution of the g-band images. In the case of NGC 1313, the FUV and g-band images had nearly identical angular resolution of 1.4\arcsec, so no additional PSF matching was required.
\section{Methodology} \label{sec4}

\subsection{Identification of star-forming clumps} \label{sec4.1}
\indent For identification of SFCs, we used the Python package \textit{Astrodendro}\footnote{\url{https://dendrograms.readthedocs.io/en/stable}} (\citealt{rosolowsky2008structural}) on our FUV images of galaxies. Astrodendro arranges the FUV emission coming from the galaxy into a hierarchical tree-like diagram called a dendrogram. The dendrogram typically consists of a main structure called the trunk - covering the entire galaxy, which splits into a large number of smaller substructures called branches, and leaves. In the context of our study, the leaves at the ends of the dendrogram tree are taken as individual SFCs, since the leaves are the most compact, densest, and unresolvable structures in the dendrogram. 

\indent The \textit{Astrodendro} package constructs the dendrogram based on three input parameters: \texttt{min\_value}, \texttt{min\_delta}, and \texttt{min\_npix}. The \texttt{min\_value} parameter sets the minimum flux level for any region to be considered as a significant structure in a dendrogram. We defined this threshold as the background flux plus three times the standard deviation (\( \text{bg} + 3\sigma \)), ensuring that structures below this level are treated as noise. The background flux and standard deviation values for the FUV images of our galaxies are calculated as the average value of the same quantities in 10 circular apertures of 1 arcminute radius each, in the sky regions, far away from the galaxy. The \texttt{min\_delta} parameter determines how distinct a peak must be from its surroundings to qualify as a separate SFC; for this, we used the standard deviation (\( 1\sigma \)) as \texttt{min\_delta}. Finally, the \texttt{min\_npix} parameter specifies the minimum number of pixels a region must contain to be identified as a structure. This was determined by examining the PSF of a point source in the UVIT FUV images (always less than 1.6\arcsec) and estimating how many pixels it spans; this turns out to be 11 pixels for NGC 5457 and 10 pixels for NGC 1313.

\indent Using these parameter choices on the FUV image, \textit{Astrodendro} identified 13,122 leaves (SFC candidates) in NGC 5457 and 4,849 leaves in NGC 1313. For each detected leaf, \textit{Astrodendro} provides a unique ID, the irregular leaf area spanned in pixels, FUV flux (in photon counts), and the central pixel coordinates (\(x_{\text{cen}}\) and \(y_{\text{cen}}\)). These physical properties of the leaves are used in subsequent analysis.

\subsection{Clump magnitudes and attenuation correction} \label{sec4.3}

\indent To perform photometry on a given SFC across different wavebands, we used circular apertures with equal area as the irregular leaf structure identified by the Astrodendro in the FUV image. These circular apertures were centered on the SFC positions provided by Astrodendro, and they enable us to determine the FUV, NUV, and g-band flux associated with each SFC. We observed that using circular apertures, as compared to the flux given by Astrodendro within the irregular leaves, results in a small amount of flux loss. However, the flux loss is within the magnitude error cuts (described later in this section) used in this paper. The choice of using circular apertures was made because Astrodendro does not allow a direct application of leaf masks onto images taken in other wavebands. Therefore, circular apertures are suitable for multi-band photometry.

The measured flux associated with any SFC is corrected for a constant sky background and the local background contribution. We elaborate on our local background subtraction method in Section \ref{sec4.4}. The final background subtracted fluxes in the FUV and NUV bands were converted to AB magnitudes using the UVIT calibration relations from \cite{tandon2020additional}, whereas the g-band fluxes, which are in nanomaggy units, were converted to AB magnitudes using the formula taken from \cite{ruiz2021characterizing}.

\indent For the FUV and NUV magnitudes of the clumps, we also estimated the magnitude errors by using error propagation on the UVIT magnitude formula. In the g-band, magnitude errors are estimated by performing aperture photometry on the g-band inverse flux maps. To ensure the reliability of observed magnitudes and ages derived from these magnitudes, 0.15 and 0.10 magnitude error cuts were applied for NGC 1313 and NGC 5457, respectively. The difference in the choice of magnitude error cuts is motivated by our need to maximize the number of SFCs for our investigation of hierarchical star formation in the later part of the paper.

\indent We corrected the observed magnitudes for Milky Way foreground, as well as galaxy-specific, constant internal dust attenuation. The foreground attenuation values are taken from the NASA dust and reddening calculator ($A_V$ = 3.1 $\times$ E (B-V); E (B-V) = 0.09 mag for NGC 1313, and 0.007 mag for NGC 5457 - measured using the recalibrated maps by \cite{Schlafly_2011}. We adopted constant internal dust attenuation \(A_V = 0.74\) mag for NGC 5457 from \citet{linden2022star} (based on spectral energy distribution fitting of star clusters) and \(A_V = 0.59 \) mag for NGC 1313 from \citet{hadfield2007survey} (based on the Balmer decrement measurement for the Wolf-Rayet stars). The corresponding $A_{\lambda}$ in the FUV, NUV, and g-band was calculated using Cardelli's extinction law \citep{cardelli1989relationship}. The total dust attenuation correction for each band was computed by adding the Milky Way and the internal attenuation contribution.

\subsection{Method for age estimation of star-forming clumps} \label{sec4.2}

\begin{table}
\centering
\caption{Key input parameters used in Starburst99}
\renewcommand{\arraystretch}{1.2}
\begin{tabular}{|c|c|}
\hline
\textbf{Parameter} & \textbf{Value} \\
\hline
Star formation type & Instantaneous \\
Initial mass function (IMF) & Kroupa [1.3, 2.3] \\
Stellar mass range & 0.1, 0.5, 120 $M_\odot$ \\
Clump mass & $10^3 M_\odot - 10^6 M_\odot$ \\
Metallicity & $Z= 0.02$ (NGC 5457) \\ & $Z=0.008$ (NGC 1313)\\
Evolutionary track & Geneva (high mass loss rate) \\
Age range & 1\textendash400 Myr \\ 
\hline
\end{tabular}\label{tab2}
\end{table}

\indent In this work, we derive ages in two similar ways - first, using FUV$-$NUV color and second, using FUV$-$g color of the SFCs. To estimate the ages of the SFCs, we used a comparison of the reddening-corrected FUV$-$NUV or FUV$-$g colors of the SFCs against the synthetic colors of star clusters derived using Starburst99 (SB99) single-aged stellar population synthesis models \citep{leitherer1999starburst99, leitherer2014effects}. We used the SB99 models to predict the UV + optical spectral evolution and the resulting color evolution for single-aged star clusters of varying masses, between ages 1 Myr to 400 Myr. We assumed instantaneous star formation and the Kroupa initial mass function \citep{2002Sci...295...82K}. Among the five metallicities available in SB99, we selected $Z=0.02$ for NGC 5457 based on metallicity estimates from \cite{vilchez2019metals} and $Z=0.008$ for NGC 1313, following the metallicity estimates by \cite{Messa_2021}, as it is the closest available value. Some of the key input parameters for generating the SB99 models are summarized in Table \ref{tab2}. 

\indent We used SB99 to generate the total stellar + nebular energy output in the form of spectral energy distributions (SEDs) for individual, unresolved star clusters of a given mass and age. The SEDs generated by SB99 lie in the wavelength range of $91\,\text{\AA}$ to $160\,\mu\mathrm{m}$ with a resolution of $\sim$10~\AA~in the UV and optical range. For our study, we focus on the wavelength range from 1250~\AA~ to 5500~\AA, which covers the wavelength range of the FUV, NUV, and g-band. The FUV, NUV, and g-band synthetic magnitudes for the simulated SFCs were calculated by convolving the SB99 spectra with the band-specific transmission rate profiles and converting the resultant fluxes into synthetic AB magnitudes. For UVIT, the effective area curves were taken from \cite{tandon2020additional}, while for the g-band, the transmission rate profile was obtained from the Spanish Virtual Observatory website\footnote{\url{http://svo2.cab.inta-csic.es/theory/fps}}. Using these magnitudes, we generated synthetic SB99 color-magnitude diagrams (CMDs) (FUV$-$NUV color vs FUV magnitude and FUV$-$g color vs FUV magnitude), corresponding to different ages and cluster masses (see Figure \ref{fig4}). From the CMDs, we can see that the SFCs get progressively redder with age, and the FUV luminosity increases with the mass of the SFC. As the different stellar mass tracks have been created in SB99 using the exact same input parameters of IMF, metallicity, and evolutionary tracks, the colour of any SFC at a given age is independent of its mass. We overplotted the observed attenuation-corrected colors and FUV magnitudes of SFCs onto these synthetic CMDs and obtained the age of each SFC by performing a linear interpolation along the color axis. In our method, the choice of the stellar mass track used for age interpolation has no impact on the derived SFC age and introduces no additional uncertainty in the age estimates.

We note that the choice of an instantaneous starburst may be an oversimplified approach, and some of these SFCs could be forming stars for a longer duration or continuously. Continuous star formation and an instantaneous burst of star formation are two extreme or limiting cases in modeling stellar populations, and they provide upper and lower limits on the age of the stellar population, respectively \citep{leitherer1999starburst99, Levesque_2013}. So, if we use a continuous star formation model, instead of an instantaneous burst, the derived ages of the SFCs would shift towards older ages (for both FUV-NUV and FUV-g based ages). In the present study, we are probing regions with typical sizes of a few tens of pc (similar to those of stellar associations and slightly larger than those of star clusters) that strongly emit in FUV. Regions of these sizes with young stars are better modeled and characterized using an instantaneous burst model \citep{2004A&A...421..887I,Pasquali_2008, 2011hsa6.conf..345S, Adamo_2017, 2019MNRAS.484.4897C, 2021MNRAS.502.1366T}, and we follow a similar approach in our study. As we have used an instantaneous burst of star-formation to characterise SFCs, the ages of the SFCs could be a lower limit. Since the main aim of this paper is to understand the hierarchical nature of star formation, which evolves with age and disperses within tens of Myr, the lower limits of ages given by the instantaneous star formation assumption are more suitable.

\subsection{Correcting the SFC fluxes for local background contribution} \label{sec4.4}
\indent As our dataset contains UV as well as optical bands, the local background contribution to the SFC fluxes in each band needs to be duly subtracted. Local background emission in galaxies is diffuse in nature and arises from the unresolved stars in the disk of the galaxy, which are not part of the SFC. In particular, the hot, evolved stellar populations could contribute to the local background emission in FUV and NUV bands. Similarly, older (age $>$1 Gyr) stellar population contributes significantly to the local background emission in g-band. Overall, the local background emission inflates the observed flux within the SFC apertures \citep{kennicutt2007star}, so it is important to subtract this contribution from the observed SFC fluxes. As the local background emission is expected to vary across the galaxy, ideally, its contribution to each SFC should be measured from annular regions around each SFC. However, this approach is difficult to implement, since our images have a relatively medium resolution of 1.6\arcsec. As we are identifying thousands of SFCs in each galaxy, at 1.6\arcsec~resolution the flux arising from neighboring SFCs will contaminate the local background measured for any given SFC in the annular region around it. 

\begin{figure}
    \centering
    \includegraphics[width=0.90\columnwidth]{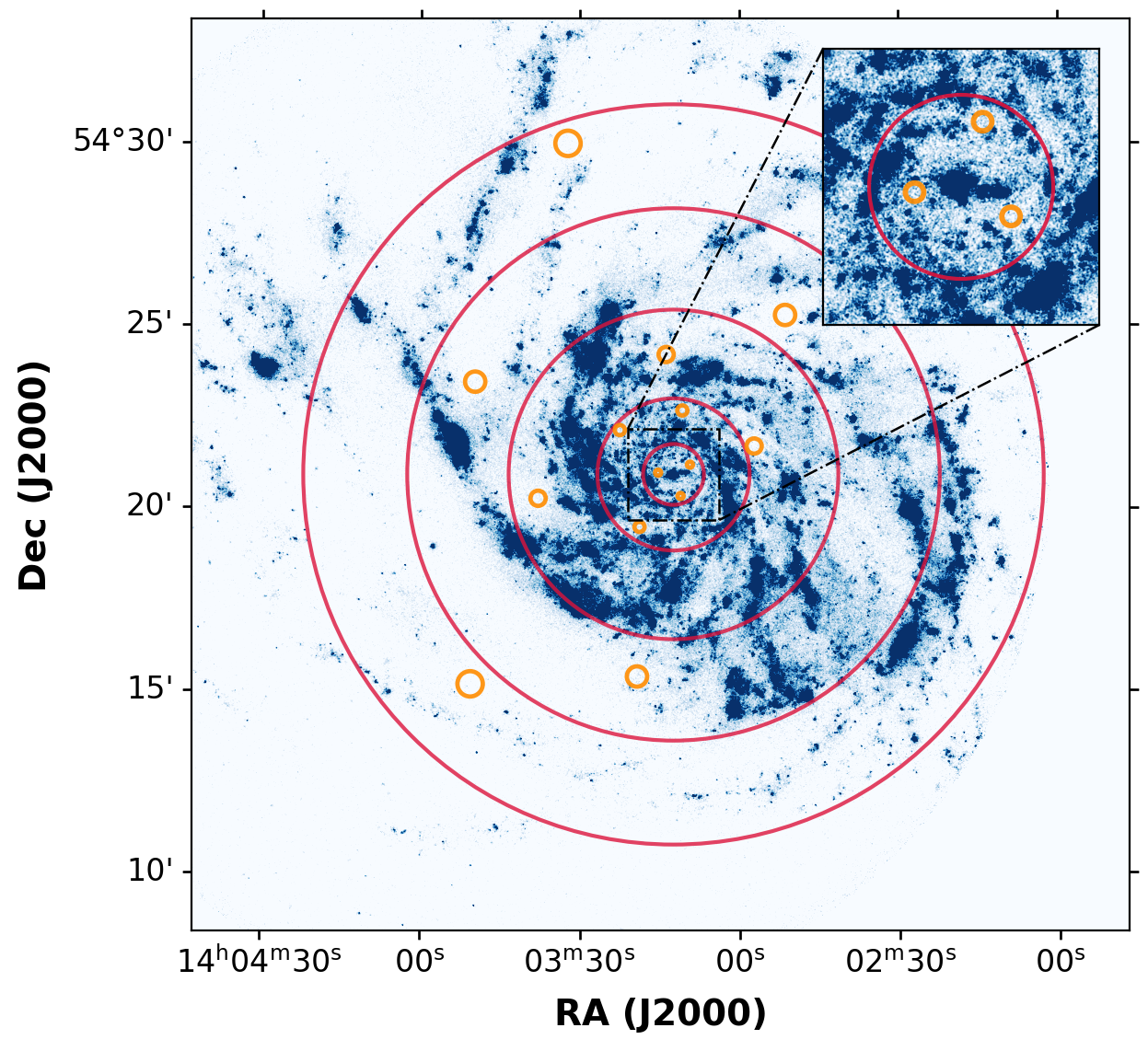} 
    \caption{A schematic illustrating our local background estimation method - as applied on the UVIT FUV image of NGC 5457. The galaxy is divided into five concentric zones (red rings). Small (orange) circular apertures of increasing radius (radii = 15, 20, 30, 40, 60) were placed in each zone for the local background estimation.}
    \label{fig2} 
\end{figure}

\begin{figure*}
    \centering
    \includegraphics[width=0.95\textwidth]{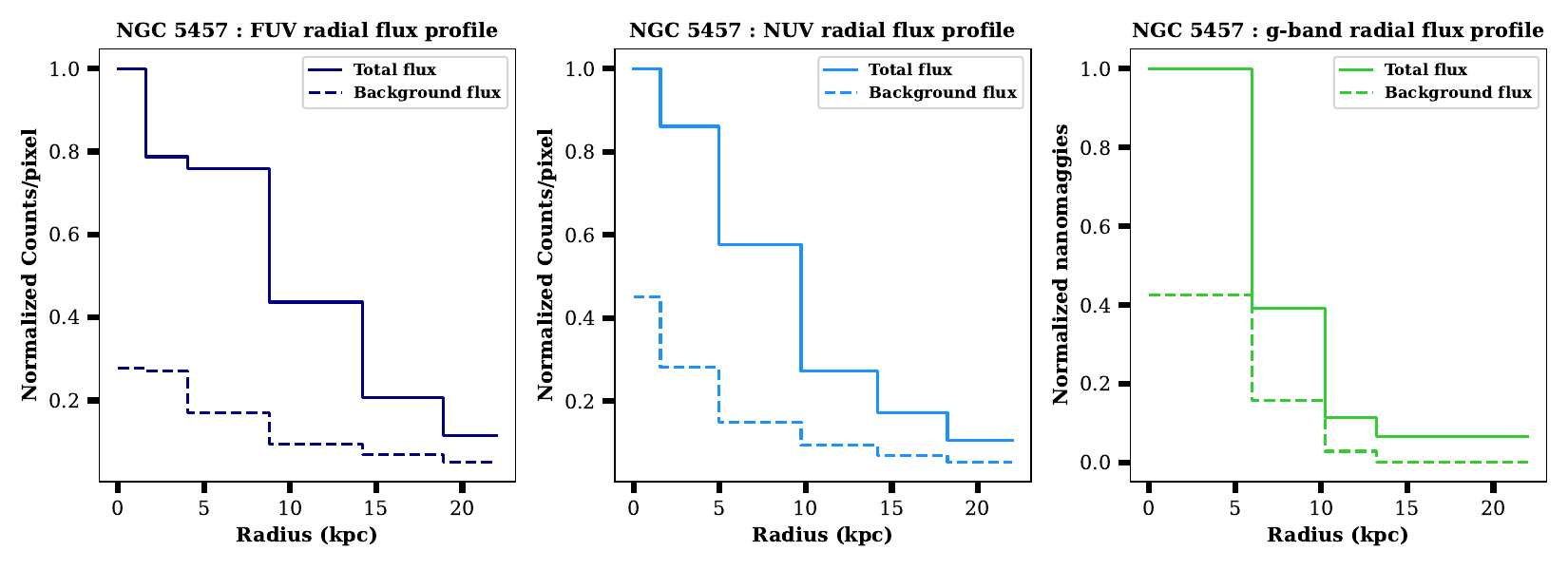}
    \caption{Left to right : The radial flux profile of NGC 5457 in FUV, NUV and g-band. Solid lines represent the normalized total flux per pixel, and dotted lines show the normalized local background flux per pixel. The normalization is performed with respect to the total flux measured at the innermost radial bin.}
    \label{fig3}
\end{figure*}

\indent In this paper, we test a pilot approach to subtract the local background contribution from the SFC fluxes. We divided the galaxy NGC 5457 into five annular radial zones. These zones are selected through visual inspection so as to delineate and cover different morphological structures of the galaxy, such as the inner or outer rings of star formation, prominent spiral arms, and bar regions of the galaxy. Within each zone, the local background flux was measured in three clump-free, circular apertures of fixed size greater than the UVIT PSF. These clump-free regions are also chosen by visual inspection, so as to probe only the diffuse background emission and to minimize contamination from star-forming regions. The local background for each zone was then taken as the average of the flux measured in these three small apertures. As it is difficult to find clump-free regions closer to the center in NGC 5457, we start with three small circular apertures of 15 pixel radius in the inner-most zone and progressively increase the aperture size to 20, 30, 40, and 60 pixels as the local background is measured in the progressively outer zones. In the outermost zone, we use the sky background as the measure of local background. This method provides a radial, zone-based estimate of the local background (see Figure \ref{fig2} for the schematic of local background measurement, and Figure \ref{fig3} for the resulting radial local background profiles). The local background emission profile derived this way falls off exponentially from the center to the outskirts of the galaxy. It is also seen that, on average, the ratio of local background emission as compared to the total emission from the galaxy increases from FUV to NUV to g-band. For NGC 1313, too, we used a similar approach to estimate the local background profiles.

\begin{figure*}[t]
    \centering
        \includegraphics[width=0.49\linewidth]{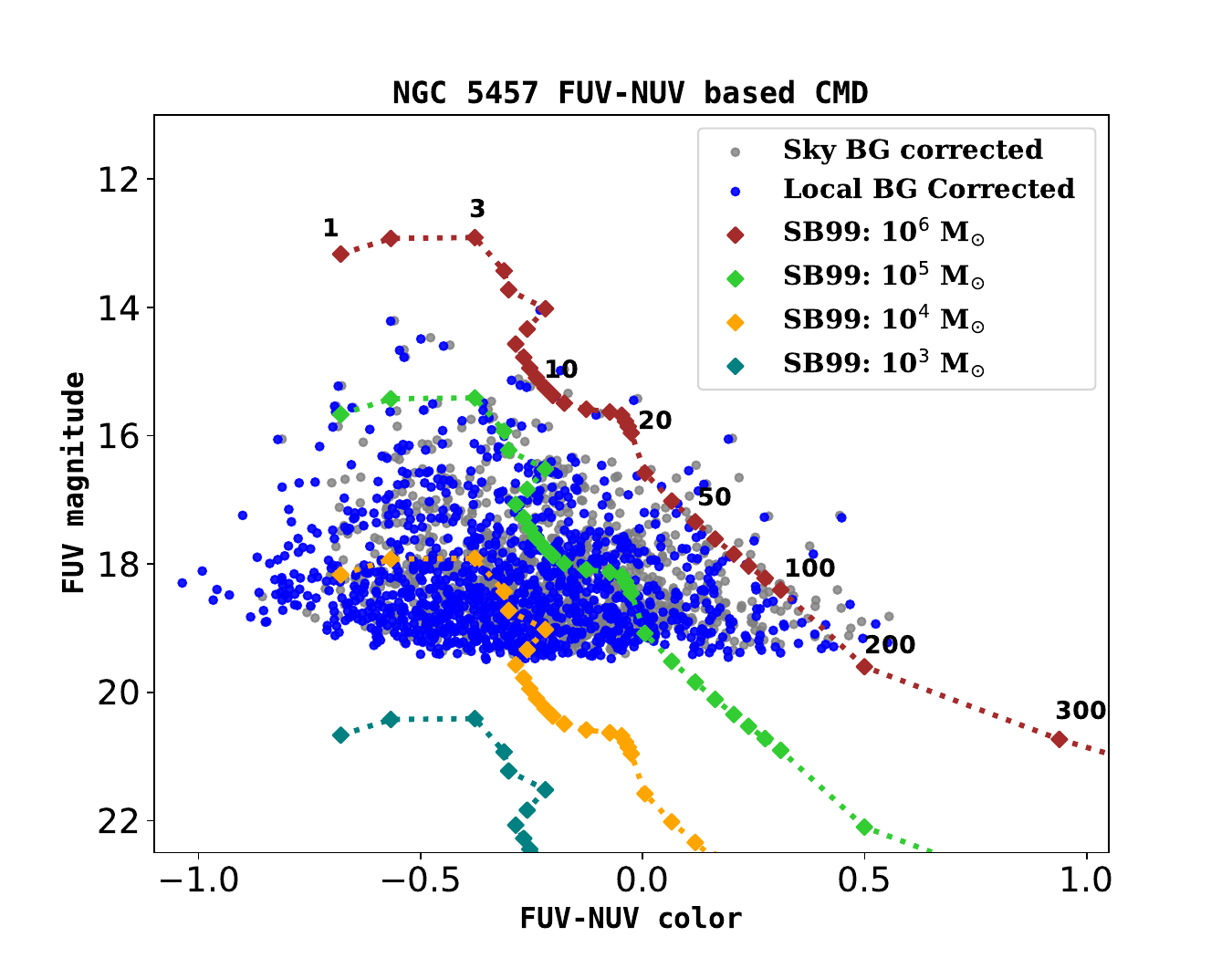}
    \hfill
        \includegraphics[width=0.46\linewidth]{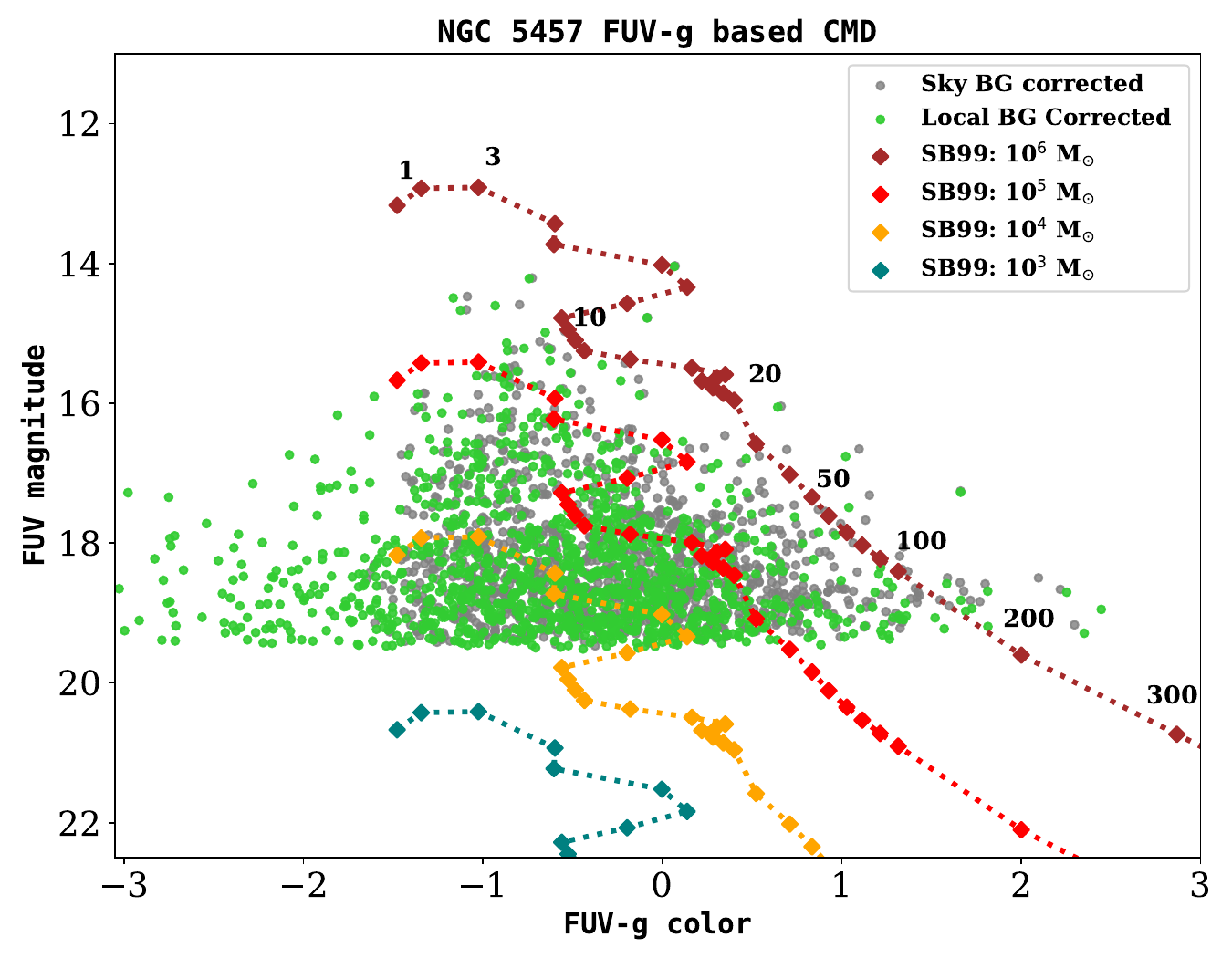}
    \caption{Left: Attenuation + local background corrected CMD (FUV$-$NUV colors vs FUV magnitude) for NGC 5457. Right: Attenuation + local background corrected CMD (FUV$-$g colors vs FUV magnitude) for NGC 5457. In both the CMDs, the blue/green points represent the local background corrected colors and magnitudes, whereas the grey points represent the sky background corrected colors and magnitudes. The four dashed tracks with diamonds represent the synthetic CMDs of star clusters of varying stellar masses. The numbers written in black on top of the tracks represent the ages of the simulated star clusters in SB99. Only the SFCs with magnitude errors cut smaller than 0.10 have been used in this analysis.}
    \label{fig4}
\end{figure*}

\indent Next, we incorporated the local background measured in each radial zone to correct the SFC fluxes. For the SFCs lying in a given radial zone, we subtracted the corresponding value of the local background from the observed SFC fluxes in FUV, NUV, and g-band, This gives us local background corrected magnitudes and colors for all the SFCs. After taking into account the dust attenuation correction for each SFC, we compared the local background corrected CMDs against the synthetic SB99 CMDs and derived the updated SFC ages (see Figure \ref{fig4}). Ultimately, we have 1122 background corrected SFCs with less than 0.10 magnitude errors in the FUV$-$NUV based CMD of NGC 5457 and 1208 background corrected SFCs with 0.10 magnitude errors in the FUV$-$g based CMD of NGC 5457. In NGC 1313, we have 1000 background corrected SFCs with 0.15 magnitude errors in the FUV$-$g based CMD. The complete catalog of SFCs containing the physical properties of SFCs in our two galaxies has been described in Appendix \ref{appdx3} and presented with the online version of this paper. The FUV, NUV, and g-band magnitude histograms for these SFCs have been presented in Appendix \ref{appdx2}. We note that for the SFCs in NGC 5457, the radii span 25.3 pc -- 178.6 pc, with a median radius of 65.3 pc. For the SFCs in NGC 1313, the radii range from 15.4 pc -- 99.5 pc, with a median radius of 25.0 pc. We use these SFC catalogs to provide a detailed discussion of the ages derived using the FUV$-$NUV and FUV$-$g color, as well as the galaxy-wide age-demographic of NGC 5457 and NGC 1313 in Section \ref{sec5}. 

\subsection{Two-point correlation function and hierarchy parameters} \label{sec4.5}
\indent The spatial distribution of young and old star-forming regions in a galaxy can be used to derive insights about the hierarchical nature of star formation \citep{grasha2017hierarchicala, menon2021dependence, shashank2025tracing, lapeer2026feast}. Since the star formation hierarchy is a direct consequence of the stars forming from hierarchically structured gas (\citealt{1981MNRAS.194..809L,1996ApJ...471..816E,lada2003embedded,elmegreen2006hierarchical,gouliermis2017hierarchical}), the youngest SFCs in a galaxy can effectively trace the hierarchical structuring of the gas from which the stars have formed. Gradually, the hierarchy of stellar structures evolves and disperses as a function of the age of the stellar population (\citealt{elmegreen2014hierarchical, grasha2017hierarchicala,menon2021dependence,shashank2025tracing, Meena_2025}). To study this effect on the hierarchical distribution of the SFCs, we used the SFC ages and the technique of spatial two-point correlation function (TPCF). The use of the TPCF technique is motivated by the fact that young star-forming regions are supposed to be spatially clustered with their neighbours, in space and time. In other words, star formation rarely happens in isolation. Rather, in a single star formation event triggered by phenomena such as cloud collisions, supernovae, feedback, spiral density waves, etc., star formation takes place simultaneously in multiple close-by locations. 

\indent In \cite{shashank2025tracing}, the TPCF method was used to derive the hierarchy parameters for NGC 5457 using sky corrected FUV$-$NUV color-based ages. In this study, we plan to compare the hierarchy parameters for NGC 5457 using FUV$-$g color-based ages against those derived in \cite{shashank2025tracing}. If the hierarchy parameters remain largely unchanged, then we can imply that the hierarchy parameters of a galaxy are insensitive to the method used for age estimation. Next, we will use the FUV$-$g color-based ages for the TPCF analysis in NGC 1313 and derive its hierarchy parameters.

\indent Here, we describe the method used to calculate the TPCF, which is based on the principle of pair counting at a given separation. TPCF is a statistical measure of the probability of finding a pair of SFCs separated by a distance 'x' as compared to two randomly distributed points in space. Interpreting the shape of the TPCF has contributed extensively to our current understanding of hierarchical star formation and the evolution of star formation hierarchies in galaxies (\citealt{zhang2001multiwavelength,sanchez2010fractal,grasha2017hierarchicalb,menon2021dependence,shashank2025tracing, Meena_2025, lapeer2026feast}). The formula used for the TPCF calculation is based on the estimator proposed by \cite{landy1993bias} called the Landy-Szalay (LS) TPCF estimator.

\begin{equation}
\begin{array}{l}
\mathrm{TPCF}(x) = 1 + w_{\mathrm{LS}}(x)
\end{array}
\end{equation}
\begin{equation}
\begin{array}{l}
w_{\mathrm{LS}}(x)=\frac{\mathrm{DD}(x)-2 \mathrm{DR}(x)+\operatorname{RR}(x)}{\operatorname{RR}(x)}
\end{array}
\end{equation}
\begin{equation}
\begin{array}{l}
\mathrm{DD}(x)=\frac{P_{\mathrm{DD}}(x)}{N_D(N_D-1)}; \mathrm{DR}(x)=\frac{P_{\mathrm{DR}}(x)}{N_D N_{R}}; \operatorname{RR}(x)=\frac{P_{\mathrm{RR}}(x)}{N_{R}\left(N_{R}-1\right)}\\
\end{array}
\end{equation} 

\indent Here, $P_{\mathrm{DD}}$, $P_{\mathrm{DR}}$ and $P_{\mathrm{RR}}$ represent data-data, data-random and random-random pairs separated by distance $x$. The total number of data points and random points considered for any TPCF analysis run are represented by $N_D$ and $N_R$, respectively. To put the formula in practice, we take different sets of SFCs binned with age (young SFCs or old SFCs) as the data distribution. Next, we populate an equal number of random points in the space spanned by the data points. Choosing each SFC as the center of our calculation exactly once, we perform the $P_{\mathrm{DD}}$, $P_{\mathrm{DR}}$ and $P_{\mathrm{RR}}$ pair counting and once all the SFCs have been considered, we finally get the TPCF as a statistical function of spatial separation 'x'. As the galaxies NGC 5457 and NGC 1313 are inclined at some non-zero angle with respect to the sky plane (see Table \ref{table1}), we de-project the SFC positions before the TPCF analysis is performed. 
We refer interested readers to \cite{shashank2025tracing} and \cite{menon2021dependence} for a more detailed explanation of the TPCF method and SFC deprojection.

\indent TPCF quantitatively analyses the distribution of SFCs, and the shape of the TPCF can help us quantify the hierarchical signatures within the distribution. The shape of the TPCF for purely hierarchical distributions should have a power-law behavior as a function of spatial scale, whereas a flat TPCF with an absolute value of 1 represents a totally random distribution. Any TPCF plot that lies between these two limits can be assumed to be a mixture of hierarchically and randomly distributed SFCs. As mentioned in Section \ref{sec1}, we expect only the youngest SFCs to show signatures of a hierarchical distribution (in the form of a power-law TPCF) since these SFCs best resemble the hierarchically structured ISM from which the stars have formed. As the SFCs get older, their hierarchical distributions gradually disperse, and they settle into the exponential disk of the galaxy. TPCF can also be used to constrain the timescale over which the hierarchy of SFCs disperses within a galaxy. We divided the SFCs into different age groups and examined how the shape of the TPCF and the hierarchical distribution of SFCs evolve with age. The different SFC age groups were made while complying with the condition that a sufficient number of SFCs need to be used in every TPCF measurement. By trial and error, we found that if the number of SFCs considered for TPCF analysis is less than $\sim$250, then the measured TPCF becomes noisy. So, for making a distinct age group of SFCs and performing the TPCF analysis on it, we take at least $\sim$250 SFCs into consideration.

\indent All of the TPCF plots generated in this paper have been fit with a broken power law of the form,

\begin{equation}
\label{eq:ModelPW}
\begin{array}{ll}
F_\mathrm{PW} (x) 
=\left\{\begin{array}{ll}
A_1 x^{\alpha_1} & :  x < l_{\rm{corr}}\\
A_2 x^{\alpha_2} & :  x > l_{\rm{corr}}
\end{array}\right.
\end{array}
\end{equation}

Here, $A_1$ and $A_2$ are constants such that $A_2 = A_1 l_{\rm{corr}}^{(\alpha_1-\alpha_2)}$; $\alpha_1$ is the power-law slope of the TPCF on small scales and it can also be used to derive the 2D fractal dimension of the stellar hierarchy as $D_2$ = 1 + $\alpha_1$. This broken power law TPCF is motivated by the physical understanding of galaxy-scale star formation hierarchies. The hierarchical signature on the SFC distribution is only expected to be observed till a maximum scale called the correlation length (at the juncture of the broken power law), and beyond the correlation length, the SFC distribution is supposed to be mostly random, showing a flat TPCF.

\section{Results} \label{sec5}

\subsection{Comparison of FUV$-$NUV and FUV$-$g based ages} \label{sec5.1}

\begin{figure*}[t]
      \centering
        \hfill
      
		\includegraphics[width=0.42\linewidth]{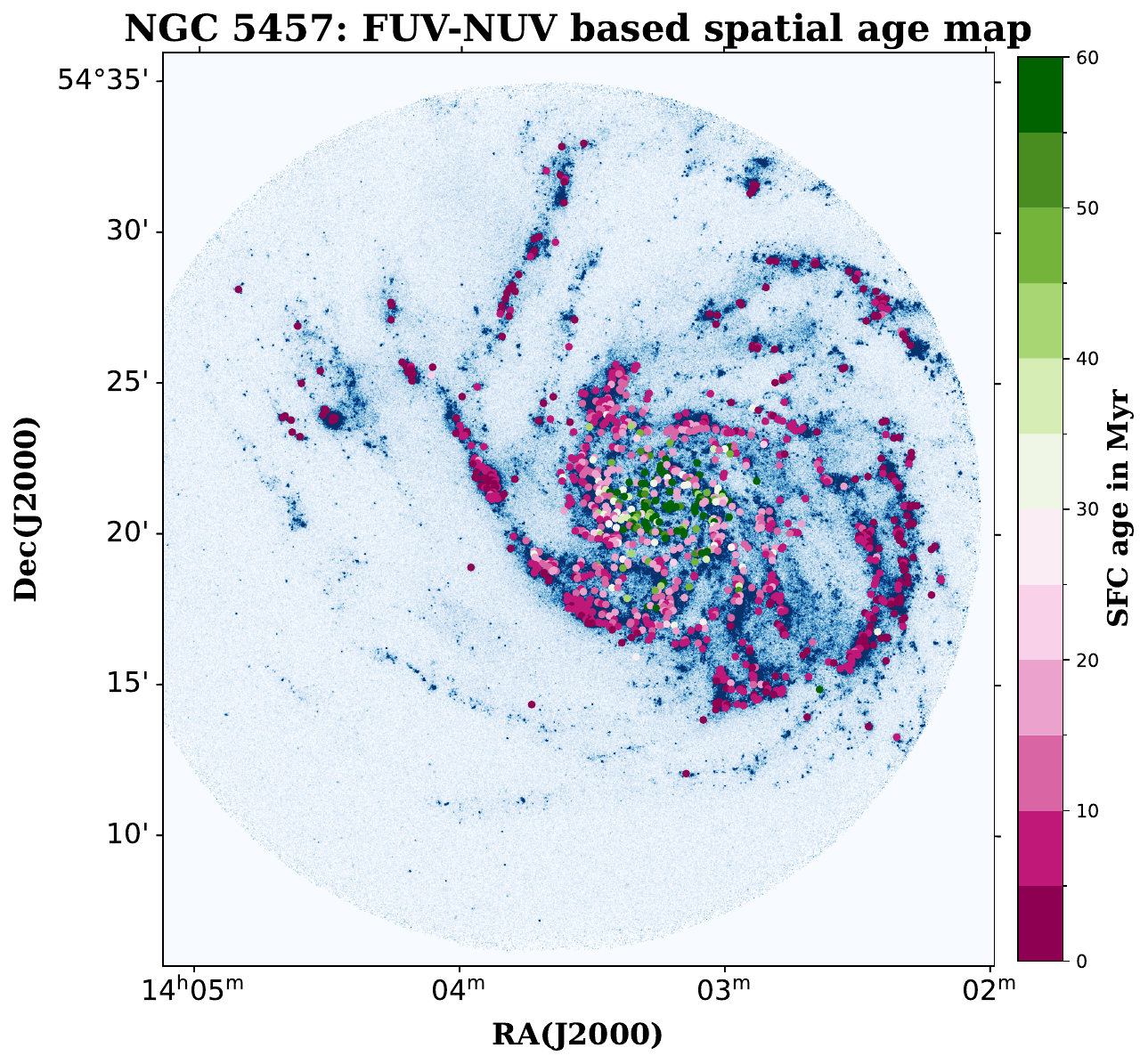}
        \hfill
		\includegraphics[width=0.42\linewidth]{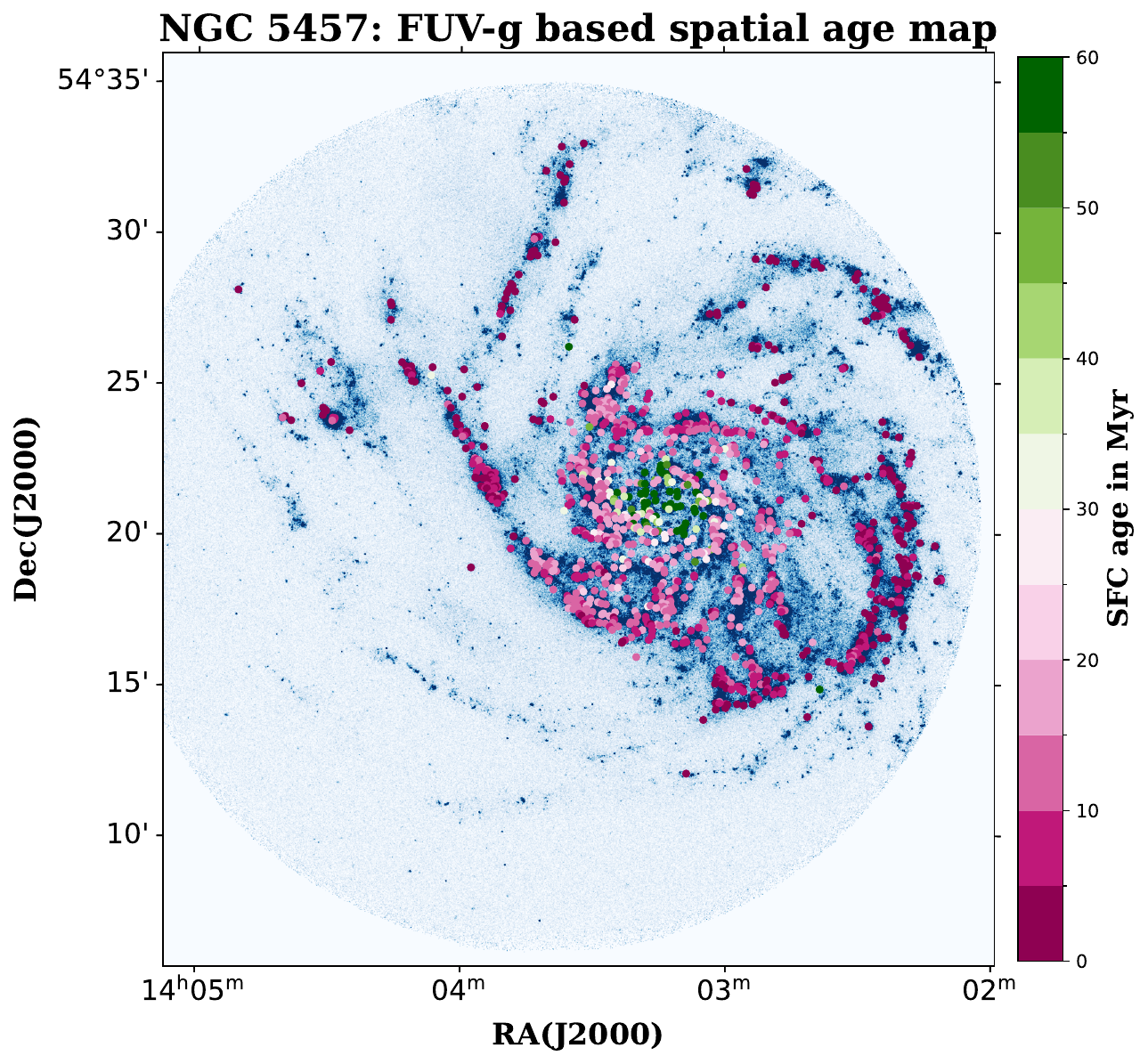}
        \hfill
        
	\vfill
        \hfill

		 \includegraphics[width=0.42\linewidth]{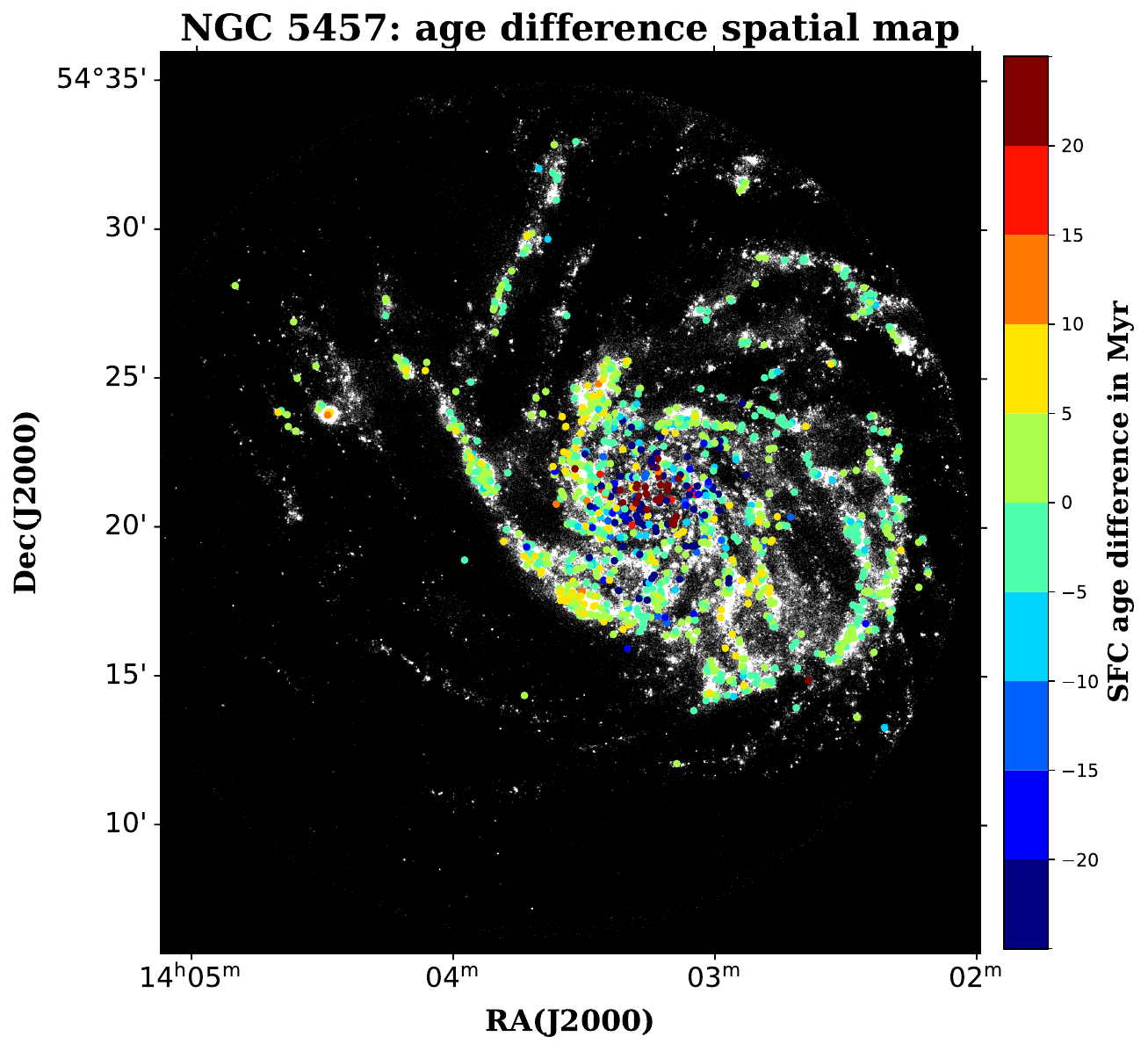}
        \hfill
		  \includegraphics[width=0.42\linewidth]{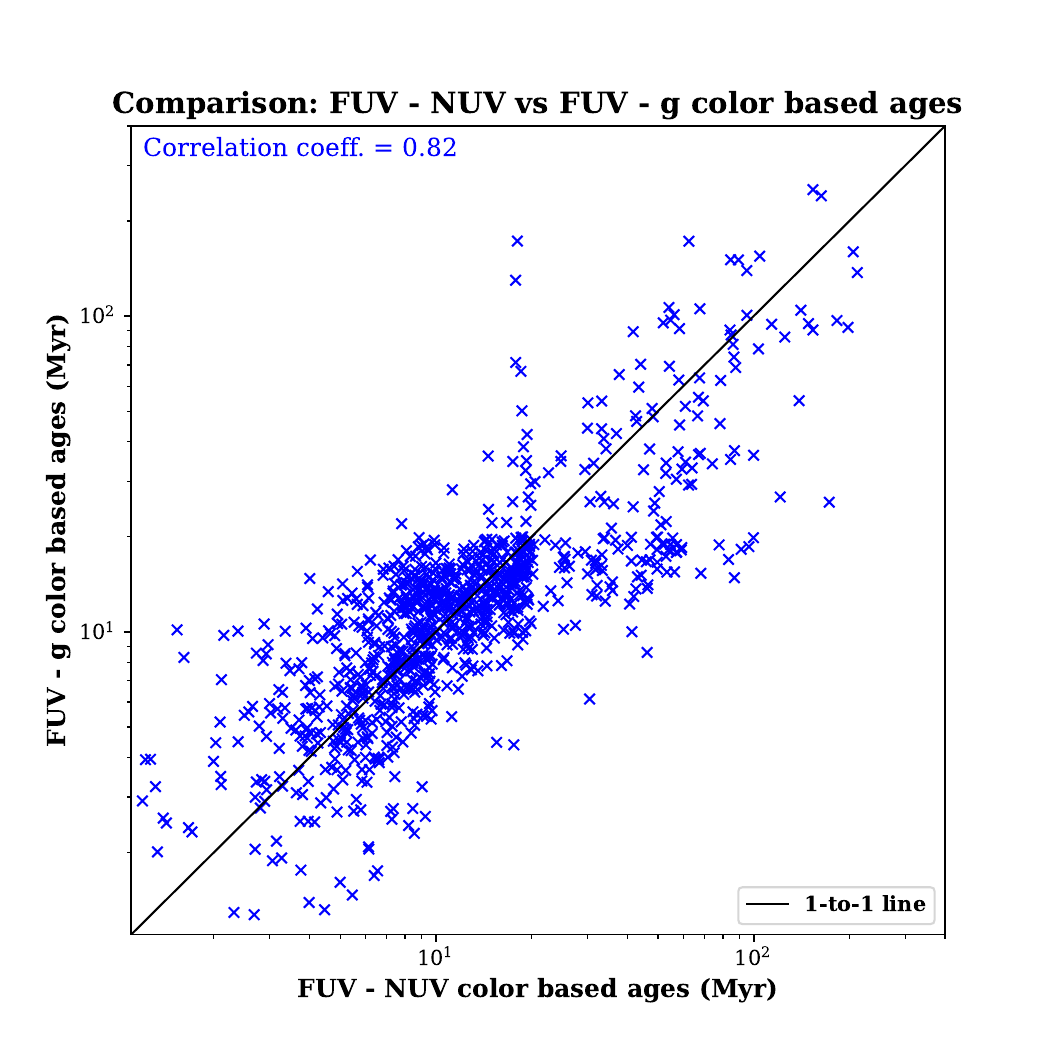}
        \hfill
          
    \caption{Top left : SFC age map of NGC 5457 derived using the local background corrected FUV$-$NUV color. Top right : SFC age map of NGC 5457 derived using the local background corrected FUV$-$g color. Bottom left : Spatial age difference map for NGC 5457 where the colorbar represents the difference between the local background corrected FUV$-$g color based ages and local background corrected FUV$-$NUV color based ages., Bottom right : A comparison between FUV$-$NUV and FUV$-$g color-based ages reveals excellent agreeement, with a Pearson correlation coefficient value of 0.82.}
  \label{fig5}
\end{figure*}

\indent As discussed in Section \ref{sec4.2}, for the age determination of the SFCs in NGC 5457, we created two types of synthetic SB99 CMDs i.e. FUV magnitude vs FUV$-$NUV color and FUV magnitude vs FUV$-$g color. We interpolated the observed attenuation + local background corrected colors against the synthetic colors from the CMDs to derive the SFC ages. This gives us FUV$-$NUV color based and FUV$-$g color based ages of the SFCs. In this section, we will compare the two different age estimates for the SFCs characterized in NGC 5457. 

\indent According to the synthetic SB99 CMD, the FUV$-$NUV color spans between -0.69 and 1.10 for ages between 1 and 300 Myrs, and the color becomes progressively redder with the age of the simulated star cluster. It is clear from Figure \ref{fig4} (left) that the majority of the observed SFCs fall within this color range, as expected from recently formed, UV emitting regions. A small fraction of SFCs ($\sim$10\%) have bluer colors than -0.69. Such SFCs are also relatively faint, so they are associated with larger magnitude and color errors. Since the SB99 spectra are quite uncertain at ages $<$1 Myr, we assume that these SFCs are younger than 1 Myr. The FUV$-$g color spans between -1.4 and 3.0 for ages between 1 and 300 Myrs (Figure \ref{fig4}, right). The FUV$-$g color range is much wider than the range of FUV$-$NUV color for the same age span. This is expected from the larger wavelength separation between the two contributing wavebands. Quite like the FUV$-$NUV CMD, most of the detected SFCs lie within an age range of 1 to 300 Myrs. The actual SFC ages are between 1-212 Myr and 1-251 Myr for NGC 5457 as derived using the FUV–NUV and FUV–g color, respectively. The median age is 9 Myr for the FUV$-$NUV based estimates and 10 Myr for the FUV$-$g based estimates. This overall agreement between the age range spanned by the SFCs, as indicated by the two CMDs, is suggestive that background corrected FUV$-$g color can serve as a reliable tracer of the SFC age, similar to the FUV$-$NUV color.

\indent The FUV$-$NUV and FUV$-$g color-based spatial age-maps (Figure \ref{fig5}, top row) suggest that on average, the inner SFCs in NGC 5457 are older and the outer SFCs are younger. By examining the age-maps and the age difference maps ((FUV$-$g color-based ages) $-$ (FUV$-$NUV color-based ages)) (Figure \ref{fig5}, bottom left), we observed good visual agreement between the ages derived from the background corrected FUV$-$NUV and FUV$-$g color. The age difference map indicates that SFCs in the outer regions show small age differences of $\pm$ 5 Myr whereas, the SFCs in the inner parts of the galaxy show age differences of $\pm$ 15 Myr. Here, we note that the SFCs (with magnitude errors less than 0.10 magnitude, as is the case with NGC 5457) in age range 1 to 20 Myr have associated mean errors of $\sim$3 Myr, and for the SFCs in age range 20 - 100 Myr, the mean age errors are $\sim$ 23Myr \citep{shashank2025tracing}. The $\pm$ 5 Myr age difference in the outer parts of the galaxy and $\pm$ 15 Myr age difference in the inner parts of the galaxy are consistent with the errors associated with the SFC ages. Overall, the inspection of age maps and age-difference map indicates a strong agreement between the SFC ages derived using the background corrected FUV$-$NUV and FUV$-$g color. 

To further quantify the agreement between the SFC ages derived using the two colors, in Figure \ref{fig5} (bottom right), we provide a one-to-one comparison of the background corrected FUV$-$NUV and FUV$-$g color-based ages. We find a statistically significant, linear relationship between the ages derived using the two different colors, as indicated by a Pearson correlation coefficient value of 0.82. Moreover, The nearly equal scatter in the positive and negative direction (as is also observed in the age-difference map) implies that choosing either the FUV$-$NUV or FUV$-$g color, as the proxy for SFC age, does not incur any systematic enhancement (or decrement) in the derived ages. This implies that using either of the FUV$-$NUV or FUV$-$g color as an age indicator leads to similar SFC ages.

\indent As shown in Figure \ref{fig3}, the local background profile of a galaxy peaks near the galaxy center and falls exponentially with galactocentric radius. Additionally, the ratio of local background contribution and the overall light profile of a galaxy depends on the wavelength of observation. The excellent agreement between the FUV$-$g and FUV$-$NUV color-based ages at larger galactocentric radii and a mixture of small positive and negative differences between the two sets of ages at smaller galactocentric radii hints that the observed age differences are perhaps caused by the differential impact of local background emission as a function of galactocentric radius and the waveband used. We aim to further refine and better optimize our methodology in future UVIT + optical-based projects. Overall, the relatively high level of convergence between the local background corrected FUV$-$NUV and FUV$-$g based ages indicates that our local background subtracted FUV–g based age estimation method is pretty well optimized at this point and can be applied to other galaxies, such as NGC 1313.

\subsection{Discussion on SFC age map of NGC 5457} \label{sec5.2}

\indent From the local background corrected age maps in Figure \ref{fig5} (top row), we can study the recent star formation history of NGC 5457 within the last $\sim$300 Myr, as indicated by its brightest SFCs. Visual inspection of the age maps of NGC 5457 indicates a nearly smooth, negative age gradient as a function of galactocentric radius - with older SFCs (green points) residing in the inner parts of the galaxy and younger SFCs (pink points) residing in the outer parts. This observation is consistent with the inside-out growth scenario of disk galaxies. The inside-out galaxy growth scenario posits that disk galaxies quickly assemble their inner, spheroidal regions early in their lifetimes so, the inner regions of the galaxy are quite old. Later, galaxies grow in size by continuously accreting neutral, star-forming gas from their surroundings \citep{1976MNRAS.176...31L,1991ApJ...379...52W,2006MNRAS.366..899N}. The slow gravitational collapse of this accreted gas sustains star formation in the galaxy outskirts and it is responsible for the observed UV emission and younger ages in the outer regions of the galaxy \citep{1976MNRAS.176...31L,2007ApJ...658.1006M, 2010ApJ...712..858G}. 

Many past studies based on GALEX FUV and NUV data also observed similar negative color-gradients (indicative of the SFC ages) in the disk galaxies such as M83 \citep{2005ApJ...619L..79T}, M31 and M33 \citep{2005ApJ...619L..67T}, NGC 5194 and NGC 5457 itself \citep{2005ApJ...619L..71B} and and other spiral galaxies \citep{2011ApJ...743..137B}. However, the GALEX resolution is 3-4 times worse than that of UVIT. Our findings suggest that even at a significantly higher angular resolution, the negative age-gradients still persist. More recently, \cite{2021MNRAS.502.5508P} also analyzed the stellar population properties of $\sim$1000 late-type galaxies using integral field unit spectroscopic data, and observed clear negative age gradients in their sample galaxies. 

The young SFCs present in the outer regions of NGC 5457 clearly indicate that the galaxy is accreting gas from the intergalactic medium, which would be consistent with its XUV disk nature and interaction history with the dwarf galaxy NGC 5474 and other companion galaxies \citep{yadav2021comparing}. It is also observed that the spiral arms of NGC 5457 are dominated by the young SFCs. This aligns with our current understanding of spiral density waves compressing molecular clouds to initiate star formation along the spiral arms \citep{1964ApJ...140..646L,cedres2013density,shabani2018searchfor}. 

In addition to inside-out disk growth, recent studies have shown that the inward migration of clumps can also contribute to the presence of older clumps in the inner regions of disk galaxies. In gas-rich, turbulent systems, massive clumps formed in the outer disk can migrate inward due to clump interactions and dynamical friction, leading to negative radial age gradients \citep{guo2018clumpy, lenkic2021giant}.

We note that residual local background contribution due to source crowding and the presence of a spheroidal bulge component close to the galaxy's center can make the SFC colors redder and the SFC ages older. This would act as a contributing factor to the observed negative radial age gradient. Additionally, variable dust attenuation in the inner vs outer regions of the galaxy can also contribute to the negative age gradient. Dust attenuation is usually higher in the inner regions of spiral galaxies, so our constant attenuation correction may be causing us to assign higher ages to the actually young SFCs in the inner regions. However, negative age gradients are also observed in other star-forming galaxies, which take local background emission and variable dust attenuation into account \citep{2020A&A...635A.177B,2021MNRAS.502.5508P,2010ApJ...712..858G,2025ApJ...990...72C}. This allows us to infer that the observed age trends observed in this paper physically exist within the galaxy.

\begin{figure*}[t]
      \centering
		\includegraphics[width=0.45\linewidth]{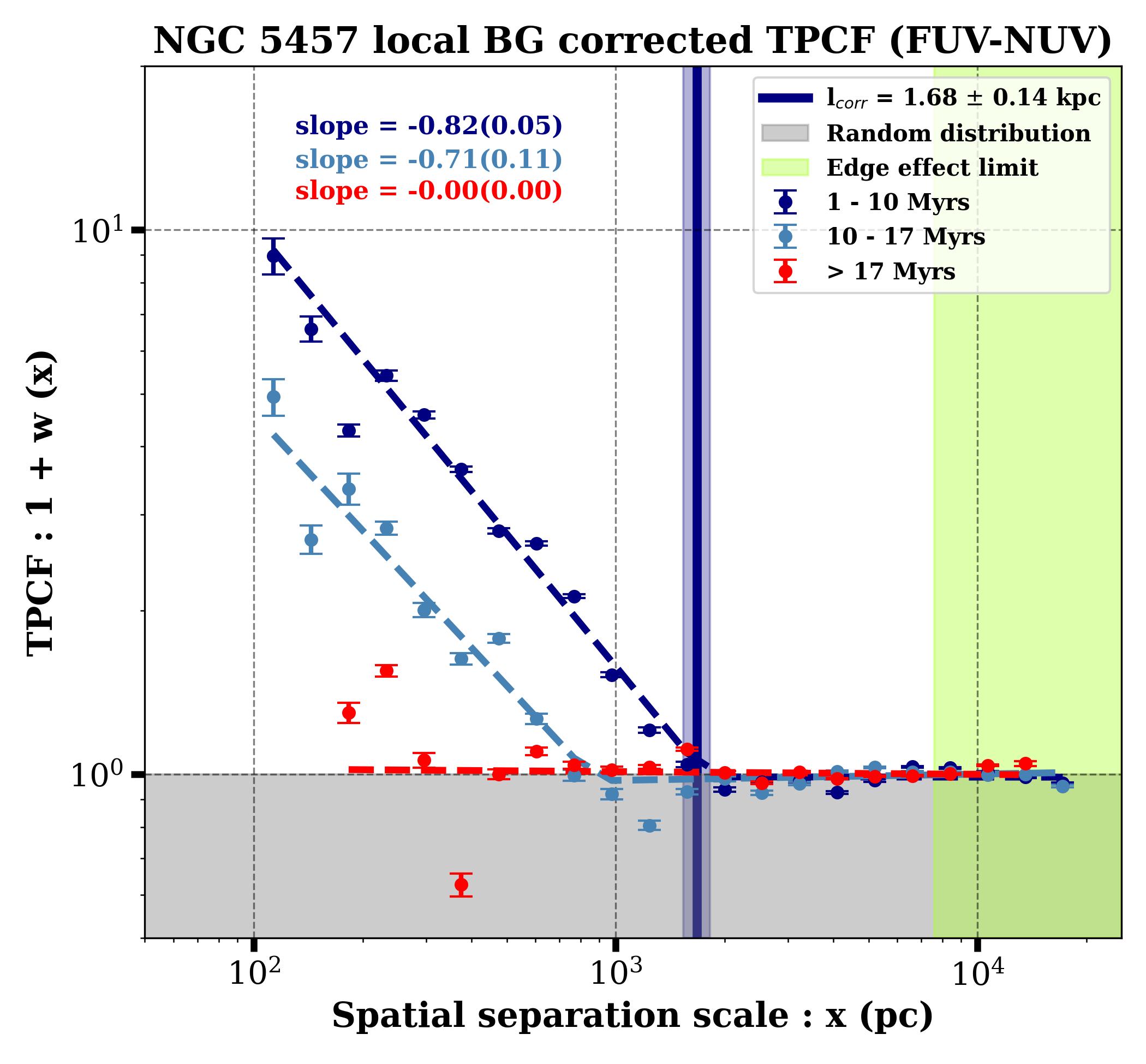}
		\label{fig:subfig1}
        \hfill
		\includegraphics[width=0.45\linewidth]{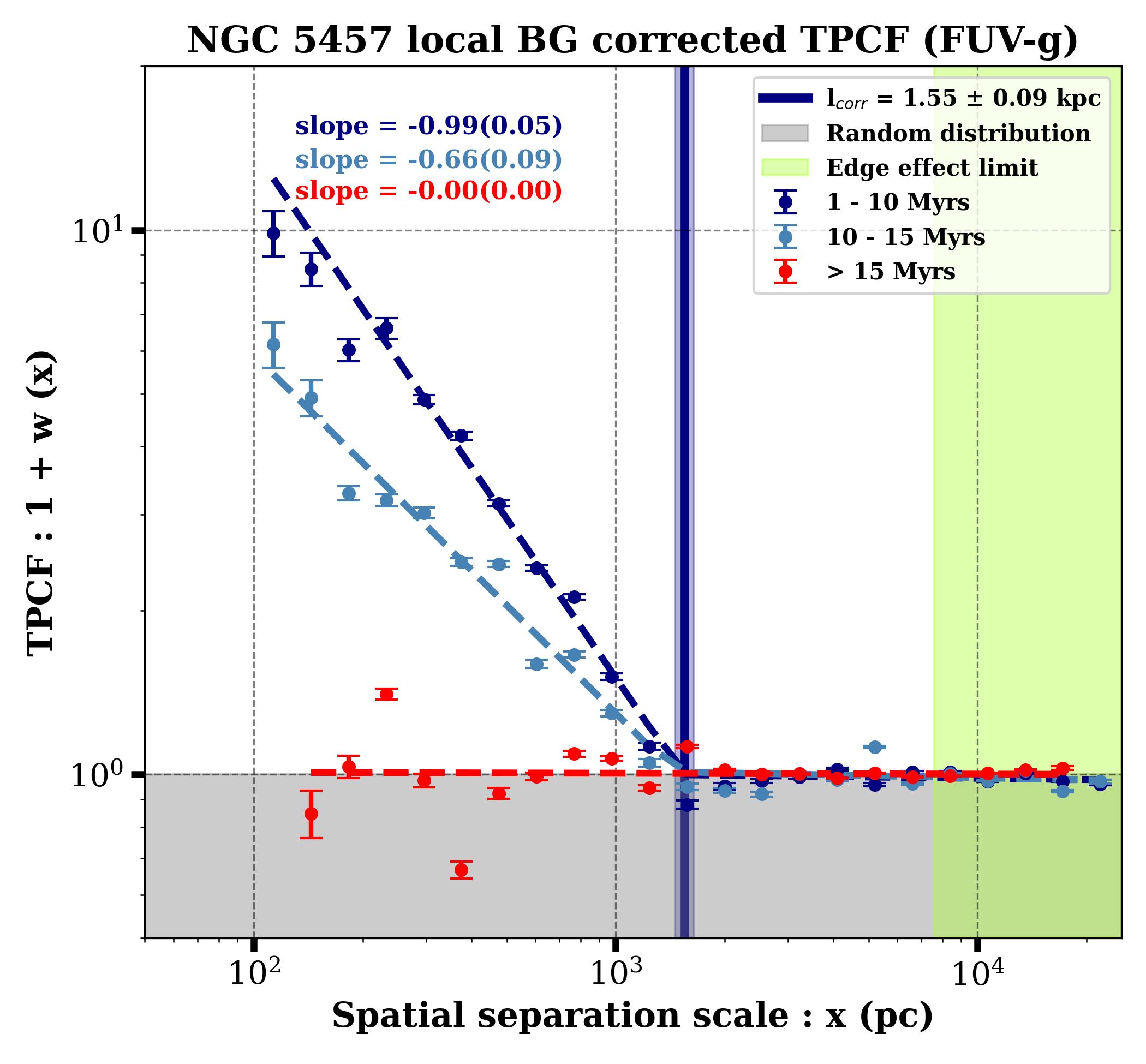}
		\label{fig:subfig2}
    \caption{Left : Results from the TPCF analysis of NGC 5457 using the local background corrected FUV$-$NUV color based ages. Right : Results from the TPCF analysis of NGC 5457 using the local background corrected FUV$-$g color based ages. In both plots, young SFCs show signs of a hierarchical distribution, as represented by the broken power-law behavior. The young SFCs reveal that the correlation length of NGC 5457 is $\sim$1.50 kpc. The older SFC TPCF implies the gradual dispersal of the hierarchical distribution of SFCs and it reveals that the star formation hierarchy disperses in NGC 5457 on scales close to $\sim$15 Myrs.}
  \label{fig6}
\end{figure*}

\subsection{Hierarchy parameters of NGC 5457 : Using FUV$-$NUV and FUV$-$g color based ages} \label{sec5.3}

\indent In Figure \ref{fig6} (left) and (right), we show the results of the TPCF analysis for the SFCs in NGC 5457 with background corrected FUV$-$NUV and FUV$-$g color-based ages, respectively. The choice of SFC age bins used in this paper is based on the requirement that, in order to robustly calculate TPCF, each SFC bin should have at least $\sim$250 SFCs \citep{shashank2025tracing}. This ensures that the TPCF plot is not too noisy and the plots can be fitted with our mathematical TPCF model. The $<$10 Myr cutoff for the youngest age bins was used in \cite{menon2021dependence} to trace the hierarchical structure expected from star-forming regions. 10 Myr is the approximate timescales in which feedback from star-forming regions disperse their parent molecular cloud (which traces the hierarchical structure of the ISM) \citep{Chevance_2020}. This allowed us to fix 1-10 Myr as our first age bin. To make our analysis robust, we also ignored SFCs with ages $<$1 Myr in our TPCF analysis (75 SFCs in FUV$-$NUV, 158 SFCs in FUV$-$g). So, our first TPCF bin corresponds to the 1-10 Myrs old SFCs. The older age bins were chosen based on the requirement of having $\sim$250 SFCs in each bin. For FUV$-$NUV based TPCF of NGC 5457, we used 1-10 Myr, 10-17 Myr, and $>$17 Myr bins containing 549, 238, and 260 SFCs, respectively. For FUV$-$g based TPCF of NGC 5457, we used 1-10 Myr, 10-15 Myr, and $>$15 Myr bins, containing 437, 306, and 307 SFCs, respectively. Unlike the FUV$–$g based analysis, we could not use the 10-15 Myr age bin for the FUV$-$NUV based analysis because of an insufficient number of SFCs (183) in that age bin. So, we updated the 10-15 Myr age bin to 10-17 Myr (containing 238 SFCs) for the FUV$-$NUV based analysis.
 
In both the plots of Figure \ref{fig6}, we observed that the 1-10 Myrs SFC population shows the broken power-law TPCF, where there exists a steep power law at the smaller scales, and on larger scales, the TPCF becomes flat. The distinct power-law behaviour of the TPCF up to the break-scale is consistent with the expectations from hierarchical distributions of SFCs in galaxies. The steep power-law behaviour is observed between scales close to the UVIT resolution ($\sim$48 pc) and up to kiloparsec scales in NGC 5457. The largest scale of this power-law TPCF behaviour, before the TPCF transitions into a flat, zero-slope regime, represents the largest scale up to which star formation in galaxies exhibits hierarchical characteristics. This scale is termed as correlation length in recent literature (\citealt{menon2021dependence, shashank2025tracing} and references therein). The FUV$-$g color-based correlation length (1.55 $\pm$ 0.09 kpc) is in good agreement with the correlation length derived using the FUV$-$NUV color-based value (1.68$\pm$0.14 kpc). The TPCF slope (-0.99 $\pm$ 0.05) derived using the FUV$-$g color based SFCs is only slightly steeper than the slope derived using FUV$-$NUV based SFCs (-0.82 $\pm$ 0.05). These mutually agreeing hierarchy parameters suggest that local background corrected FUV$-$g colors are also equally as effective as the FUV$-$NUV colors in constraining the hierarchy parameters of a galaxy. This match provides additional validity to our SFC ages derived using FUV$-$g colors and shows that our method is effective in characterizing the young SFC population of the galaxy. Therefore, with the encouraging results derived for NGC 5457 using local background corrected FUV$-$g based ages, we can apply the same method to study the star formation hierarchy in NGC 1313.

As we take older SFC sets into consideration for TPCF analysis (10-17 Myr and $>$17 Myr SFCs in the FUV$-$NUV based ages, 10-15 Myr, $>$15 Myr SFCs in the FUV$-$g based ages), the slope of the TPCF plot and the absolute amplitude of the TPCF are found to be lower than the values observed for the young SFCs (1-10 Myrs). This behavior of older SFC TPCF is consistent with a progressively dispersing distribution of SFCs as a function of age, governed by the large-scale galaxy dynamics. This also implies that these older SFCs are not suitable for measuring the true hierarchy parameters of any galaxy, since they represent a deviation away from the initial SFC hierarchy. We note that in \cite{shashank2025tracing}, a slope limit of -0.2 was taken as an indication of a near-complete dispersal of the SFC hierarchy. For the FUV$-$NUV (FUV$-$g) color-based TPCF, we observed a TPCF slope of -0.71 $\pm$ 0.11 (-0.66 $\pm$ 0.09) for the 10 - 17 Myr (10 - 15 Myr) SFCs, which is consistent with a significantly hierarchical distribution. However, for both the FUV$-$NUV and FUV$-$g color-based TPCF, we observed a near-zero TPCF slope for the $>$17 Myr and $>$15 Myr SFCs, respectively. The near-zero slope for the oldest SFC age bins can also be affected by the really noisy nature of the TPCF, due to small number statistics. However, the fact that the absolute correlation value for this age bin is always close to 1 (as expected from a completely random distribution) allows us to infer that the SFC hierarchy disperses by 17 Myrs (15 Myrs) in NGC 5457 - as derived based on the FUV$–$NUV (FUV$-$g) color-based ages. We note that as we are limited by small number of SFCs in the oldest age bin, our hierarchy dispersal timescales may represent lower limits. 

These behaviors shown by the older than 10 Myr SFCs signify that the SFC distribution loses its memory of ever being a hierarchical distribution and is rapidly dispersing. The 15-17 Myr hierarchy dispersal timescale of NGC 5457 matches quite well with the 20 Myr dispersal timescale derived in \cite{shashank2025tracing}. Many studies suggest that eventually, when the hierarchies are completely dispersed, the SFCs merge into the exponential disk of the galaxy (\citealt{menon2021dependence,grasha2017hierarchicala,shashank2025tracing}). We also observed that there is good qualitative agreement between the TPCF behavior of old SFCs in NGC 5457 irrespective of whether the ages are derived with FUV$-$NUV or FUV$-$g color.

\indent We note that the hierarchy parameters of NGC 5457, reported in this study (correlation length = 1.68$\pm$0.14 kpc and TPCF slope= $-$0.82$\pm$0.05) using the local background corrected FUV$-$NUV color-based ages, are slightly lower than those derived in \cite{shashank2025tracing}  (correlation length = 1.90$\pm$0.10 kpc and slope= $-$0.95$\pm$0.05), without local background subtracted FUV$-$NUV color-based ages. However, the values are comparable to each other and lie in the same ballpark. The observed differences could be attributed to the small variations in the age estimates due to the local background correction performed in this study and the resulting difference in the spatial distribution of young SFCs used to measure TPCF. Additionally, we note that in \cite{shashank2025tracing}, $\sim$1600 sky-corrected, less than 0.1 magnitude error-cut SFCs were identified in NGC 5457. But, in this work, we only have $\sim$1100 SFCs with less than 0.1 magnitude error-cut, after local+sky background subtraction. The loss of SFCs in this paper is caused by the local background subtraction leading to a reduction in SFC fluxes and enhancement in the SFC magnitude errors. Subsequently, we get lesser number of SFCs with 0.1 magnitude error, post local background subtraction. 

We also refer the interested readers to Appendix \ref{appdx1}, where we perform a bootstrap analysis to investigate the impact of using SFCs with different magnitude error cuts, (which impacts the uncertainties in age estimates) on the hierarchy parameters.

\begin{figure*}[t]
   \centering
    \includegraphics[width=0.45\linewidth]{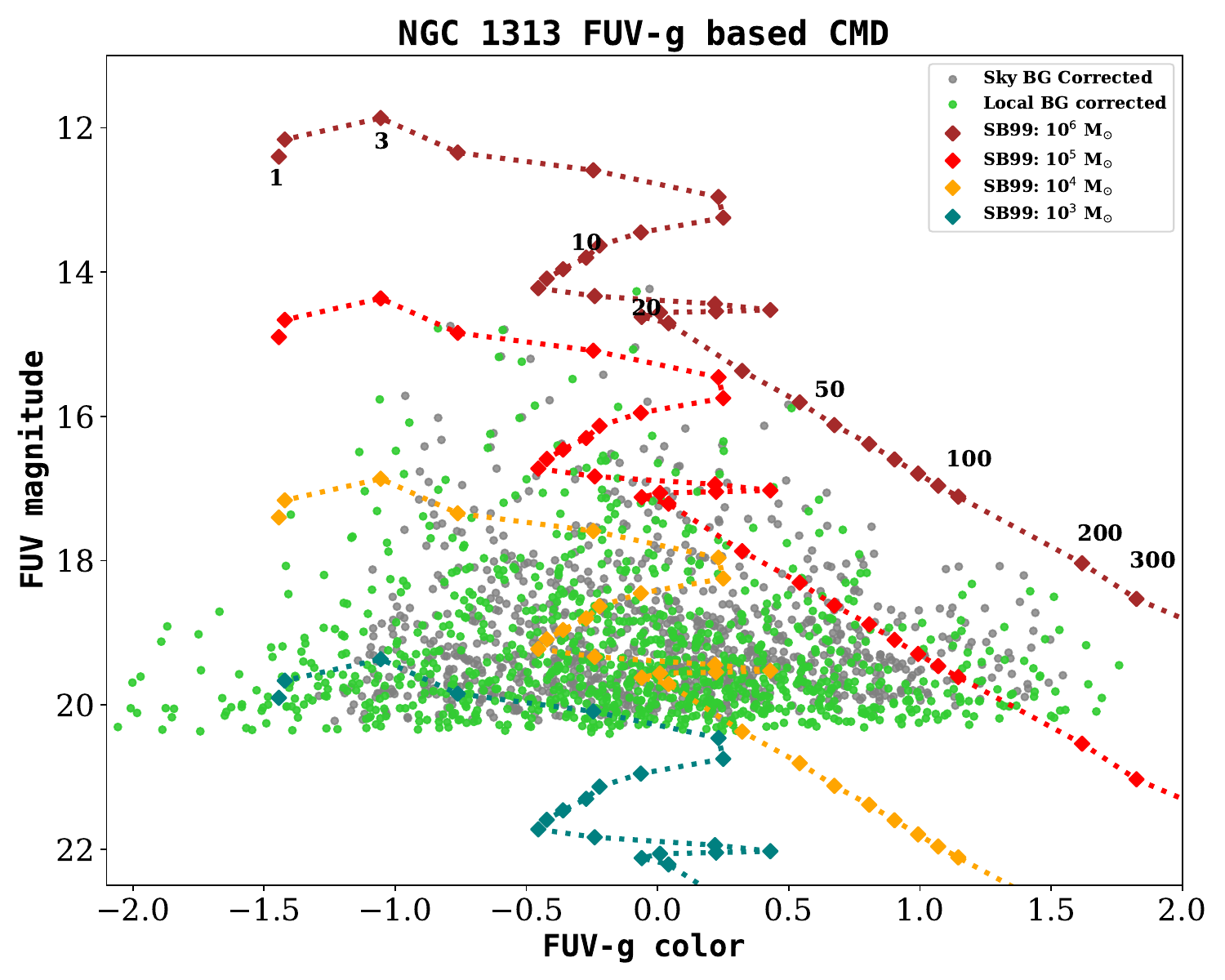}
    \includegraphics[width=0.42\linewidth]{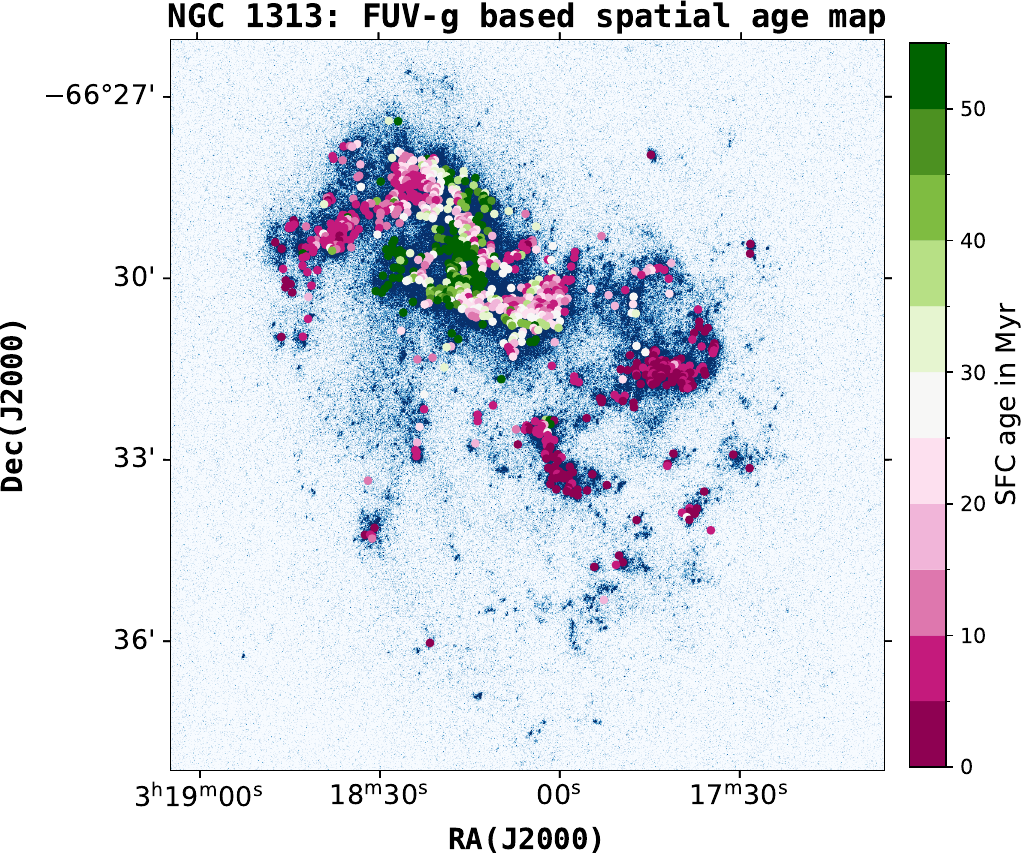}
    \caption{Left : Attenuation + local background corrected CMD (FUV$-$g colors vs FUV magnitude) for NGC 1313. The green/gray points represent the local/sky background corrected colors and magnitudes. The four dashed tracks with diamonds represent the synthetic CMDs of star clusters of varying stellar masses. The numbers written in black on top of the tracks represent the ages of the simulated star clusters in SB99. Right : SFC age map of NGC 1313 derived using the sky background corrected FUV$-$g color. Only the SFCs with magnitude errors cut smaller than 0.15 have been used in this analysis.}
    \label{fig7}
\end{figure*}

\subsection{Local background corrected age map of NGC 1313} \label{sec5.4}
\indent Having tested and verified our age-estimation methodology on NGC 5457, we applied the same methods on NGC 1313. Since the UVIT NUV observations for NGC 1313 are not available, we derived the attenuation and local background corrected FUV–g color-based ages for its SFCs. The age-map of NGC 1313 (see Figure \ref{fig7}) reveals a concentration of young SFCs (pink points) in the outer regions of NGC 1313 and an abundance of old SFCs (green points) closer to the center. The SFC age distribution spans from 1 to 268 Myr, with a median age of 15 Myr. Similar to NGC 5457, these age trends are consistent with the inside-out growth scenario of disk galaxies \citep{1976MNRAS.176...31L,1991ApJ...379...52W,2006MNRAS.366..899N}. However, the disk structure of NGC 1313 is not very clear rather, it has a peculiar morphology. The main body of NGC 1313 is a barred spiral galaxy, with distorted spiral arms, patchy star formation in the south-west direction, and diffuse shells encompassing the entire galaxy \citep{messa2021looking,2022AJ....164...89H}. \cite{1994MNRAS.269.1025P} used the HI observations of NGC 1313 to infer that the galaxy must have gone through a recent interaction event. Subsequent studies by \cite{2012MNRAS.423..213S,2022AJ....164...89H,2024ApJ...964...13F} have taken the increased star formation activity in the south-west part as compared to the rest of the galaxy to confirm the interaction status of the galaxy. These studies suggest that perhaps an interaction with a minor satellite galaxy triggered a recent, local burst in the south-western part of the galaxy. The complex morphology of NGC 1313 may therefore be attributed to its past interaction history.

The age map of NGC 1313 suggest that the SFC population in the bar region of NGC 1313 (the bar is aligned roughly in the north-south direction) appears relatively older than the SFCs present in its two spiral arms. However, not all the SFCs in the bar region are old and intermediate age SFCs (white points) are also present in the bar region. The young + intermediate age SFCs in NGC 1313 resemble a horizontally flipped S-shaped pattern. Similar S-shaped pattern, induced by past interactions, was also observed in the distribution of young stars (ages $<$30 Myr) in the interacting, irregular galaxy NGC 4449 \citep{2025ApJ...990...72C}. Similar to the findings by \cite{2012MNRAS.423..213S,2022AJ....164...89H,2024ApJ...964...13F}, the SFC population in the south-east direction of the galaxy, which is disjoint from the main body, appears to be quite young in our age map. In the LEGUS covered, inner parts of the galaxy, our age demographic is also qualitatively consistent with the star cluster age map shown in \cite{menon2021dependence}, which proves the reliability of our SFC ages. 

\subsection{Hierarchy parameters of NGC 1313} \label{sec5.5}
\begin{figure}
      \centering
		\includegraphics[width=0.90\linewidth]{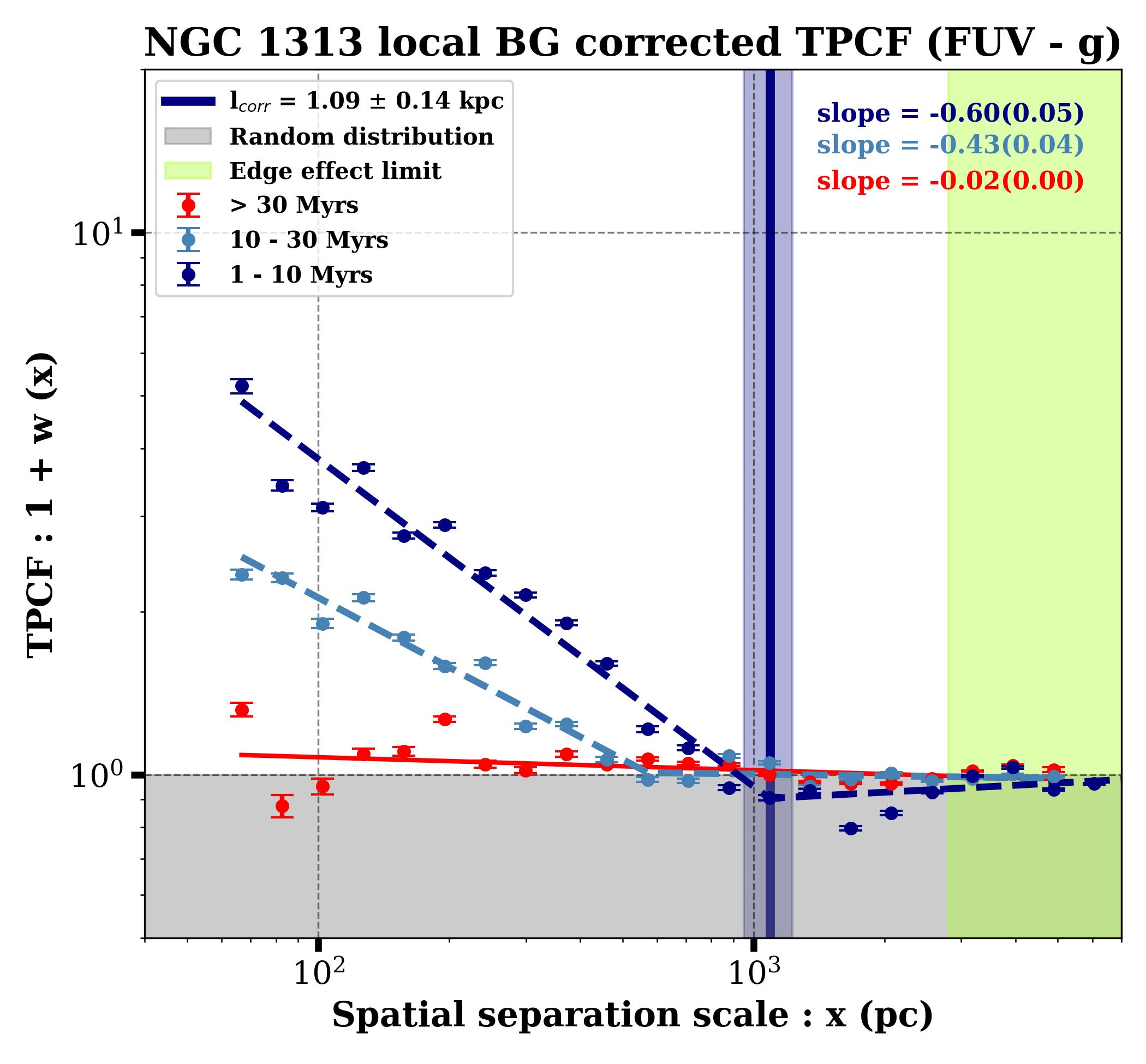}
		\label{fig:subfig1}
    \caption{Results from the TPCF analysis of NGC 1313 using the local background corrected FUV$-$g color based ages. Similar to NGC 5457, the TPCF analysis suggest that the young SFCs show a strongly hierarchical distribution. The correlation length of NGC 1313 is found to be $\sim$1.09 kpc, which agrees very well with the results from \cite{menon2021dependence}. The older SFC TPCF implies that the star formation hierarchy disperses in NGC 1313 on timescales close to $\sim$30 Myrs.}
  \label{fig8}
\end{figure}

\begin{table*}[t]
\centering
\caption{A comparison of the hierarchy parameters of NGC 5457 and NGC 1313 as measured in different studies (In the references, M21 =\cite{menon2021dependence} and GS25 = \cite{shashank2025tracing}}
\begin{tabular}{cccccccc}
\hline
Galaxy & BG correction & Color used & Correlation length & TPCF slope & Primary data, galaxy coverage & Reference \\
 & method & & (kpc) & Young SFCs & & & \\\hline
NGC 5457 & sky BG & FUV$-$NUV & 1.90 $\pm$ 0.10 kpc & -1.04 $\pm$ 0.12 & UVIT, full coverage & GS25 \\
NGC 5457 & local BG & FUV$-$NUV & 1.68 $\pm$ 0.14 kpc & -0.82 $\pm$ 0.05 & UVIT, full coverage & This work \\
NGC 5457 & local BG & FUV$-$g & 1.55 $\pm$ 0.09 kpc & -0.99 $\pm$ 0.05 & UVIT, full coverage & This work \\
NGC 5457 & local BG & SED fitting & 0.45 $\pm$ 0.18 kpc & -0.60 $\pm$ 0.14 & HST, NW part coverage & M21 \\\hline
% NGC 1313 & sky BG & FUV$-$g & 1.66 $\pm$ 0.20 kpc & -0.54 $\pm$ 0.03 & UVIT & full & This work \\
NGC 1313 & local BG & FUV$-$g & 1.09 $\pm$ 0.14 kpc & -0.60 $\pm$ 0.05 & UVIT, full coverage & This work \\
NGC 1313 & local BG & SED fitting & $>$0.96 kpc & -0.60 $\pm$ 0.03 & HST, main body coverage & M21 \\\hline
\end{tabular}
\end{table*}

\indent In Figure \ref{fig8}, we show the results for TPCF analysis of the SFCs in NGC 1313 with local background corrected FUV$-$g color-based ages. We have a total of 1000 SFCs with 0.15 magnitude errors after the local background subtraction. We divided these SFCs into 1-10 Myrs, 10-30 Myrs,  and $>$30 Myrs age bins. These bins contain 357, 336, and 264 SFCs, respectively. This division aims to reveal both the initial conditions and the age evolution of the star formation hierarchy in the galaxy. Additionally, the age bins are chosen in such a way that each bin contains at least 250 SFCs, which aids in a robust, less-noisy TPCF measurement. Similar to the results from NGC 5457 and the expectation from a truly hierarchical distribution, we observed that the youngest age bin (1-10 Myr) SFCs show the steepest slope and highest amplitude of TPCF. When this TPCF plot is fitted with a broken power-law, the largest scale of star formation hierarchy i.e the correlation length of NGC 1313 is revealed to be 1.09 $\pm$ 0.14 kpc. This correlation length measurement is in agreement with the results from \cite{menon2021dependence}, where the lower limit for the correlation length was estimated to be 960 pc. The slope of our TPCF is -0.60 $\pm$ 0.05 also greatly agrees with the slope of -0.60 $\pm$ 0.03 derived in \cite{menon2021dependence}. We believe that the excellent match between our hierarchy parameters with those derived in \cite{menon2021dependence} is possibly because, unlike their very small coverage of NGC 5457, a significantly large part of NGC 1313 (43\% by area), including its bar and spiral arm regions is covered in their observations. For older SFC age bins considered for NGC 1313, i.e, 10-30 Myrs and $>$30 Myr SFCs, we observe a progressively diminishing amplitude of the TPCF and a general decreasing trend of the TPCF slope. This is consistent with the physical scenario where star formation hierarchies are dispersed with age. The extremely shallow slope of -0.02 $\pm$ 0.00 for the older than 30 Myrs SFC population suggests that the star formation hierarchy in NGC 1313 is dispersed by 30 Myrs timescales. These timescales are consistent with the similar timescales derived in \cite{grasha2017hierarchicalb}, \cite{menon2021dependence}, and \cite{shashank2025tracing}. 

\section{Summary and Future work} \label{sec6}
\indent In this study, based on the UVIT and legacy survey observations of NGC 5457, we present a methodology for estimating the ages of SFCs using FUV$-$g color and validate it using the FUV$-$NUV color-based age estimates. One of our aims with this paper was to evaluate whether the FUV–g color-based ages can serve as a reliable alternative to the conventional FUV–NUV color-based ages, as was used by past UVIT papers in the literature \citep{Mondal_2021, 10.1093/mnras/stac2285}. This study is performed considering the non-operational state of UVIT's NUV telescope and the wealth of archival FUV-only UVIT data, which can be utilized to study hierarchical star formation in nearby galaxies. We successfully validated our FUV$-$g color-based age estimation approach with NGC 5457 and then applied it on NGC 1313. 

\indent Using UV data from UVIT and optical g-band data from the legacy survey, we identified star-forming clumps (SFCs) in our galaxies NGC 5457 and NGC 1313 using the astrodendro source detection package. We performed aperture photometry on the SFCs across all bands using photutils and determined the multi-band magnitudes and colors of the SFCs. We used SB99 \citep{leitherer1999starburst99} stellar population synthesis model to derive the SFC ages based on the observed, dust attenuation-corrected FUV$-$NUV and FUV$-$g colors. In the process, we implemented a radial zone-based method to estimate local background flux (diffuse emission arising from stars that are not physically part of the SFCs) and subtracted it from the SFC fluxes in order to derive accurate SFC ages. Next, we used the age maps derived with the local background corrected, FUV$-$NUV and/or FUV$-$g color-based ages to study the age distribution of SFCs within our galaxies.

\indent We also utilized the SFC ages for the investigation of hierarchical star formation in our 2 galaxies. The hierarchical properties of NGC 5457 were previously derived in \cite{menon2021dependence} with extremely small galaxy coverage using the HST observations and in \cite{shashank2025tracing} with full galaxy coverage using the UVIT. NGC 1313 was studied in \cite{menon2021dependence} with partial galaxy coverage, due to which its hierarchy parameters were poorly constrained. We used two-point correlation function to constrain the hierarchical properties of the young and old SFCs in the two galaxies, and to study the initial configuration and the age evolution of stellar hierarchy in our galaxies. A comparison of our derived hierarchy parameters for the galaxies with those derived in the aforementioned studies allowed us to derive insights about using FUV + optical (g-band) data for SFC age determination, the effect of performing local background correction, and the impact of different amounts of galaxy coverage on the hierarchy parameters. The major takeaway points from our work are as follows.

\begin{enumerate}
    \item With the example of NGC 5457, which has FUV, NUV and g band data available, we demonstrated that the derived ages of the SFCs using the local background corrected, FUV$-$NUV color and FUV$-$g color are in excellent agreement with each other. This suggests that in the absence of NUV data, the FUV$-$g color can act as an equally reliable age tracer as the FUV$-$NUV color.
    \item We estimated the ages of the SFCs in NGC 1313 using the attenuation and local background subtracted FUV$-$g colors. 
    \item Our TPCF analysis for studying hierarchical star formation in our sample galaxies revealed that the youngest SFC population shows clear and strong signatures of being part of a hierarchy, whereas older SFCs show signatures of a gradually dispersing hierarchical distribution. 
    \item  The young SFCs are spatially correlated with each other on scales ranging from a few tens of parsecs up to a few kiloparsecs. The largest scale of this correlation is called the correlation length and it is revealed to be 1.55 $\pm$ 0.09  kpc and 1.09 $\pm$ 0.14 kpc for NGC 5457 (Figure \ref{fig6}) and NGC 1313 (Figure \ref{fig8}), respectively. 
    \item The full galaxy coverage of NGC 1313 using UVIT, allowed us to derive the correlation length of NGC 1313 for the first time.
    \item Based on the FUV$-$g based TPCF analysis, we found that the hierarchy dispersal timescale of NGC 5457 and NGC 1313 is $\sim$15 Myr and $\sim$30 Myr, respectively.
    \item Our TPCF analysis for the two galaxies demonstrates the importance of a telescope like UVIT in constraining the hierarchy parameters of nearby galaxies, aided by the full galaxy coverage it provides due to its 28\arcmin~FoV. 
\end{enumerate}

\indent 
Building on this paper, our local background corrected, FUV$-$g color-based methodology can be employed for the exploration of hierarchical star formation in an even larger sample of nearby galaxies. This can lead to important insights about the dependence of hierarchy parameters on host galaxy properties \citep{grasha2017hierarchicala,menon2021dependence,shashank2025tracing}. In this context, 17 galaxies with FUV and NUV observations available from the UVIT have already been studied in Shashank et al. a,b (submitted, in prep., respectively). They characterized $\sim$25000 SFCs in their galaxies of diverse morphologies and explored the properties of the star formation hierarchy in these galaxies. For a bigger sample of approximately 30 UVIT-observed galaxies with existing FUV-only data, we can use the FUV$-$g color-based method outlined in this paper for studying hierarchical star formation. In these future studies, we aim to also refine our local background subtraction technique and potentially improve our age estimates.\\

\section*{Data Availability} \label{sec7}
The AstroSat UVIT data used in this paper are publicly available and can be accessed at \url{https://astrobrowse.issdc.gov.in/astro_archive/archive/Home.jsp} using the proposal IDs G05\_233 (PI: askpati) for NGC 5457 and A07\_149 (PI: chayan) for NGC 1313. Additionally, the science-ready Legacy Survey images utilized in this study are available at \url{https://www.legacysurvey.org/viewer}.
\\

\begin{acknowledgements}
      The authors thank the anonymous referee for their valuable comments on the manuscript. SA thanks the Indian Institute of Astrophysics (IIA) for providing the opportunity to work under the Visiting Student Program and also wishes to thank Rakshit, Himanshu, Renu, and Amalchand for their valuable help during the initial stages of this work. SS acknowledges the support from the Science and Engineering Research Board of India through POWER grant(SPG/2021/002672) and support from the Alexander von Humboldt Foundation. SHM acknowledges the Centre for Computational Astrophysics (CCA) at the Flatiron Institute, a division of the Simons Foundation. CM acknowledges support from the National Science and Technology Council, Taiwan (grant NSTC 112-2112-M-001-027-MY3) and the Academia Sinica Investigator award (grant AS-IA-112-M04). This work made use of the following Python packages for data analysis and plotting: Astropy(\citealt{2013A&A...558A..33A, 2018AJ....156..123A}), Matplotlib(\citealt{Hunter07}), NumPy(\citealt{2020Natur.585..357H}), SciPy(\citealt{2020NatMe..17..261V}), and Photutils(\citealt{2024zndo..10967176B}). We also acknowledge the use of the software SAOImage DS9(\citealt{2003ASPC..295..489J}).
\end{acknowledgements}

\bibliography{ref.bib}{}
\bibliographystyle{aasjournal}

\appendix

\section{Bootstrap analysis to study the effect of magnitude error cuts on the TPCF}
\label{appdx1}
The hierarchy parameters are highly dependent on the selection of young SFCs. We have used the colors of the SFCs to estimate their ages, and larger magnitude errors will lead to larger uncertainties in the age estimates. In \cite{shashank2025tracing}, it was demonstrated that small age errors (resulting from small magnitude errors) lead to accurate measurements of the TPCF and the associated hierarchy parameters. They showed that if SFCs with large age/magnitude errors are used in TPCF analysis, it flattens the TPCF slope, and the correlation length is found to be higher. This happens because SFCs with larger age errors can cause a significant number of older than 10 Myr SFCs to be classified as younger than 10 Myr. The inclusion of these SFCs in the less than 10 Myr SFCs can dilute the correlation, leading to a shallower slope and a higher correlation length measurement. Hence, the SFCs with larger magnitude errors are excluded in this paper in order to obtain better estimates of SFC ages. In this paper, we optimized the magnitude error cut by also ensuring that a sufficient number of SFCs are available for TPCF analysis. So, 0.10 and 0.15 magnitude error SFCs were used in the TPCF analysis of NGC 5457 and NGC 1313 respectively.

Including all the SFCs, without applying any error cut, and using bootstrap analysis can help to better understand the impact of uncertainties in age estimates on the final hierarchy parameters. For the bootstrapping analysis, we considered all the SFCs with magnitude errors less than 0.2 mag. We exclude the SFCs with errors more than 0.2 mag. This cut is essential for ensuring that the detected clumps are significant detections (above 5 times the noise) and are not spurious sources. For all the SFCs with mag error less than 0.2 mag, the ages are estimated.

\begin{figure*}
      \centering
		\includegraphics[width=0.45\linewidth]{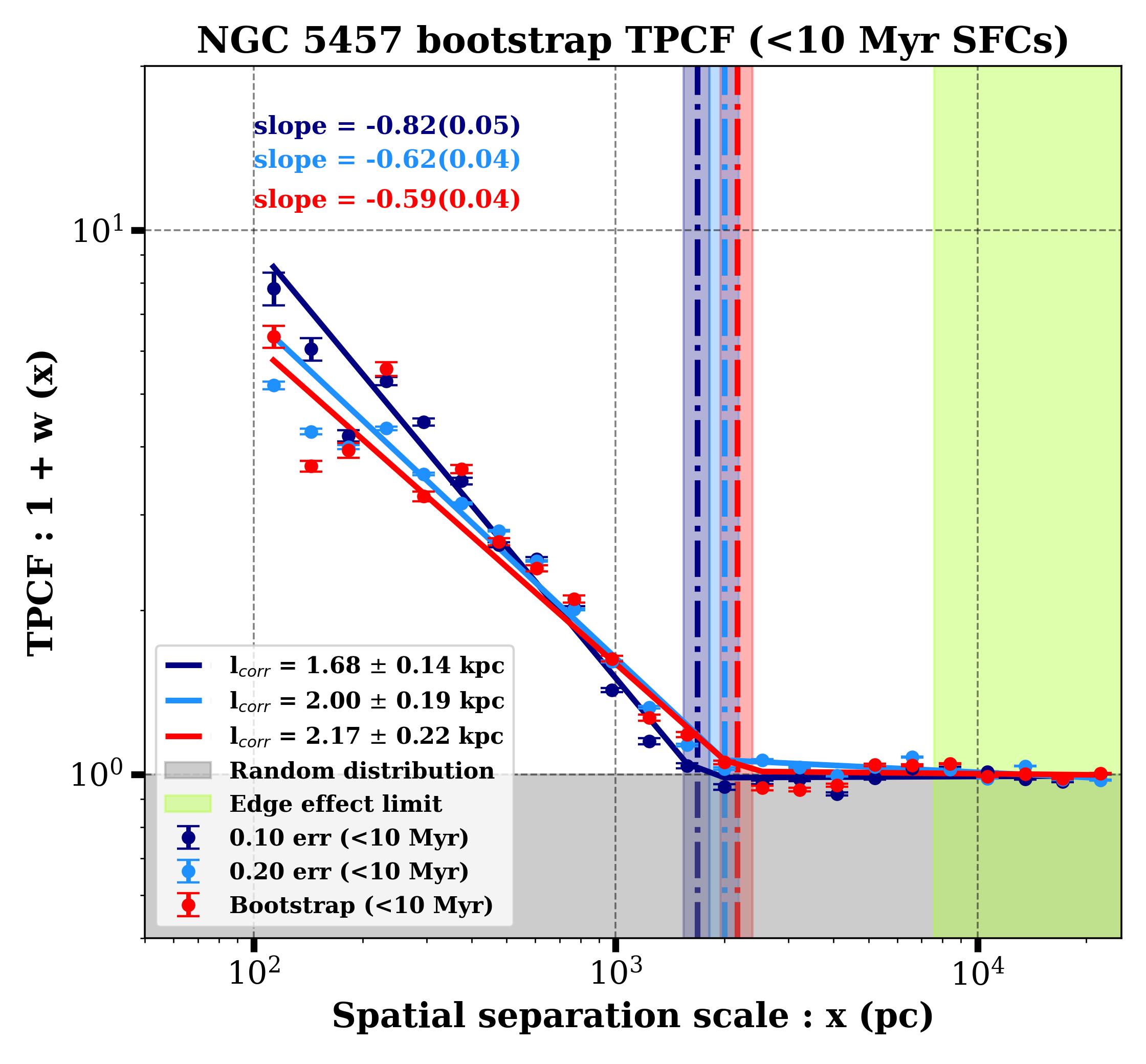}
		\label{fig:subfig1}
    \caption{Results from the bootstrap TPCF analysis of NGC 5457 using FUV$-$NUV based ages (see Appendix \ref{appdx1} for the interpretation)}.
  \label{fig9}
\end{figure*}

As the hierarchy parameters are derived using young SFCs, in Figure \ref{fig9} we show the TPCF analysis of $<$10Myr SFCs (FUV$-$NUV based ages) of NGC 5457 in three ways. The dark blue points/line shows the TPCF as measured with the young SFCs having magnitude errors less than 0.1 mag ($\sim$549 in number; slope $\sim$-0.82, correlation length $\sim$1.68 kpc) - same plot as the one shown in Figure \ref{fig6} in the paper. The light blue points/line represent the TPCF as measured with the young SFCs having magnitude errors less than 0.2 mag ($\sim$1425 in number). The TPCF slope is shallower, and the correlation length is higher with these SFCs (slope $\sim$-0.63, correlation length $\sim$2.00 kpc). The red points/line represent the TPCF measured for SFCs with ages younger than 10 Myr using the bootstrap method. To implement the bootstrap method, we took 100 iterations of selecting 549 SFCs out of the 1425 total SFCs (with less than 0.2 magnitude error), with replacement. We also generated $\sim$549 random points in the same footprint as the SFCs to act as the random sample in our bootstrap TPCF analysis. By performing the TPCF analysis on the 549 points selected in each of the 100 iterations, coupled with the $\sim$549 random points, we obtained the red points belonging to the bootstrap TPCF plot. The resulting slope is $-$0.59, and the measured correlation length is $\sim$2.17 kpc, comparable to that of the values estimated using all the SFCs with mag errors less than 0.2 mag. These results are qualitatively similar to what is demonstrated in \cite{shashank2025tracing}.

This analysis suggests that SFCs with the small magnitude errors, which in turn translate to smaller age uncertainties, are more suitable for deriving accurate hierarchy parameters. Overall, we conclude that whenever there are enough SFCs with less magnitude/age uncertainties to perform TPCF analysis, it is preferred to use them for better estimation of hierarchy parameters 

\section{Local background corrected magnitude histogram}
\label{appdx2}

\begin{figure}
      \centering
		\includegraphics[width=0.48\linewidth]{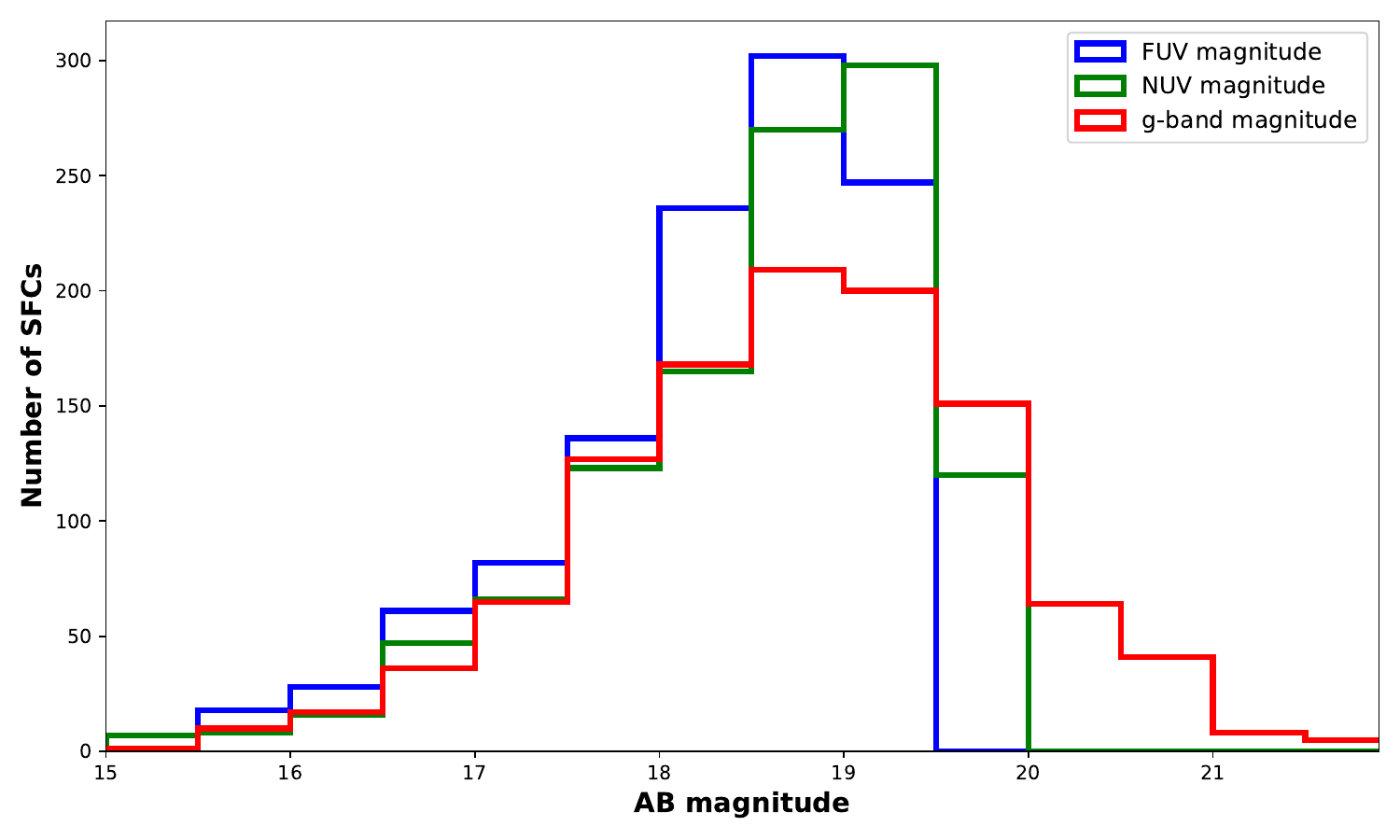}
        \includegraphics[width=0.483\linewidth]{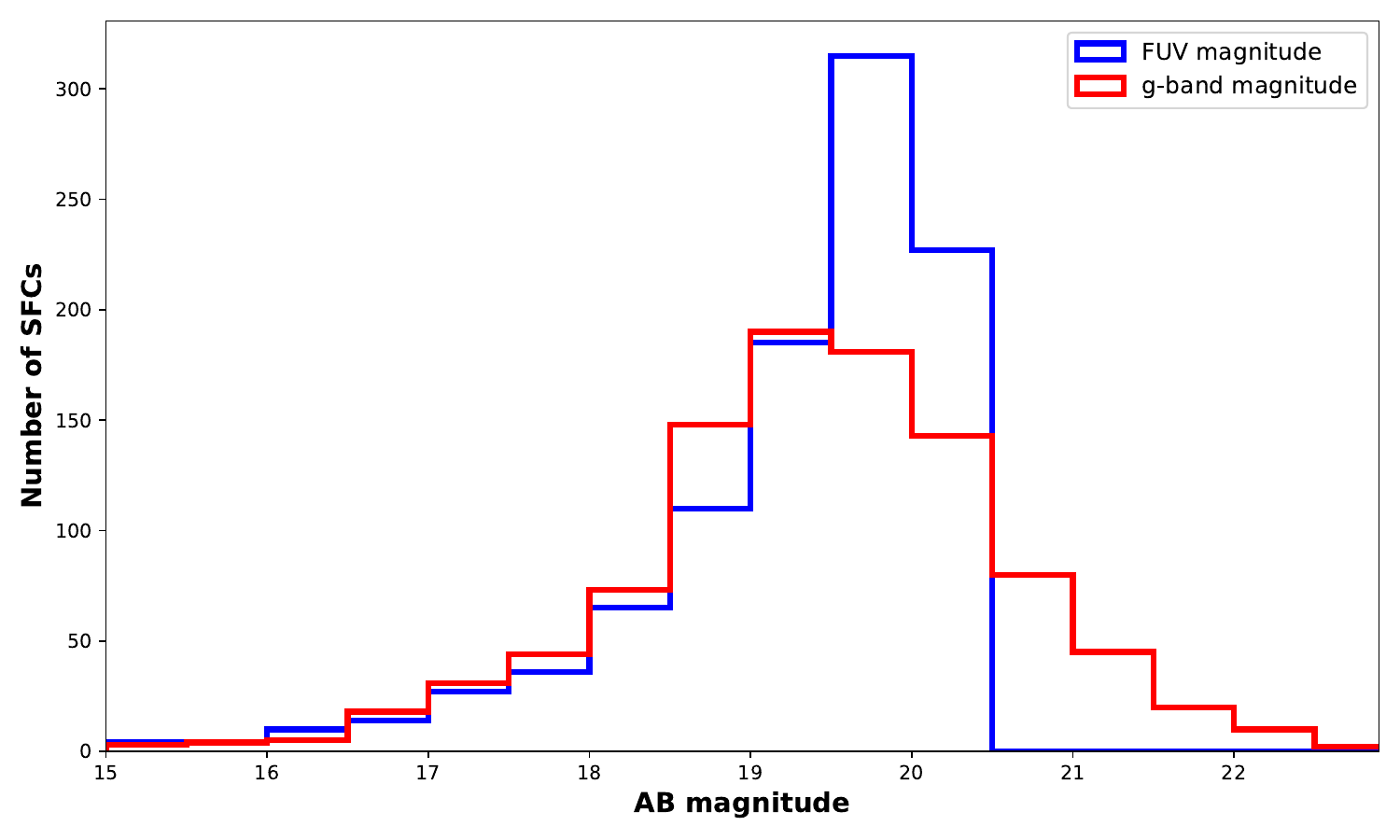}
    \caption{Left: FUV, NUV, and g-band magnitude histogram of NGC 5457. Right: FUV and g-band magnitude histogram of NGC 1313.}
  \label{fig10}
\end{figure}

\section{SFC catalogs for NGC 5457 and NGC 1313}
\label{appdx3}

\begin{table}
\centering
\caption{SFC age catalog for NGC 5457 based on FUV-NUV color}
\label{tab:catalog5457NUV}
\begin{tabular}{cccccccccc}
\hline
Id. & Area & RA & Dec & FUV mag & NUV mag & FUV mag err & NUV mag err & FUV$-$NUV & Age  \\
& (pix) & (deg) & (deg) & (mag) & (mag) & (mag) & (mag) & (mag) & (Myr)  \\
(1) & (2) & (3) & (4) & (5) & (6) & (7) & (8) & (9) & (10) \\
\hline
49 & 75 & 210.8837 &	54.5500	& 18.39 &	19.04 &	0.06 & 0.07 &	-0.64&	1.73 \\
56 &	84 &	210.9047 &	54.5482 &	18.30 &	19.10 &	0.05 & 	0.07 &	-0.79 &	1.00 \\
204 &	129 &	210.9194 &	54.5349 &	18.49 &	18.89 &	0.06 &	0.07 &	-0.40 &	6.78 \\
237 &	131	& 210.9061 &	54.5325 &	18.81 &	19.42 &	0.07 &	0.09 &	-0.60 &	2.49  \\
266 &	162 &	210.9012 &	54.5304 &	16.70 &	17.13 &	0.02 &	0.02 &	-0.43 &	6.17 \\
\hline
\end{tabular}
\tablecomments{This table shows a representative portion of the full catalog containing 1122 SFCs.(1): Id assigned to SFCs by Astrodendro. (2): SFC area spanned in pixels. (3) and (4): right ascension and declination (reported to four decimal places). (5) and (6): attenuation + local background corrected FUV and NUV magnitudes of the SFCs, and their associated errors are listed in (7) and (8), respectively. (9): FUV$-$NUV color. (10): SFC age based on FUV$-$NUV color. The full catalog is available in machine-readable format. All tabulated values are reported to two decimal places, while the full catalog provides the exact derived values. }
\end{table}

\begin{table}
\centering
\caption{SFC age catalog for NGC 5457 based on FUV$-$g color}
\label{tab:catalog5457g}
\begin{tabular}{cccccccccc}
\hline
Id. & Area & RA & Dec & FUV mag & g-band mag & FUV mag err & g-band mag err & FUV$-$g & Age \\
 & (pix) & (deg) & (deg) & (mag) & (mag) & (mag) & (mag) & (mag) & (Myr)  \\
(1) & (2) & (3) & (4) & (5) & (6) & (7) & (8) & (9) & (10) \\\hline
49 &	75 &	210.8837 &	54.5500 &	18.39 &	20.46 &	0.06 &	0.01 &	-2.06 &	1.00 \\
56 & 84 &	210.9047 &	54.5482 &	18.30 &	20.64 &	0.05 &	0.01 &	-2.34 &	1.00 \\
143 &	114	& 210.8972 &	54.5385 &	18.90 & 22.41 &	0.08 &	0.08 &	-3.50 &	1.00 \\
198 &	152 &	210.7288 &	54.5359 &	18.55 &	22.10 &	0.07 &	0.09 &	-3.55 &	1.00 \\
204 & 129 &	210.9194 &	54.5349 &	18.49 &	20.65 &	0.06 &	0.01 &	-2.16  &	1.00 \\
\\
\hline
\end{tabular}
\tablecomments{This table shows a representative portion of the full catalog containing 1208 SFCs.The column descriptions are the same as in Table \ref{tab:catalog5457NUV}, except for Columns (6), (8), (9), and (10), which correspond to the g-band magnitudes, its associated error, FUV$-$g color, and the SFC ages derived from the FUV$-$g color, respectively. }
\end{table}

\begin{table}[t]
\centering
\caption{SFC age catalog for NGC 1313 based on FUV-g color}
\label{tab:sfc_catalog_1313}
\begin{tabular}{cccccccccc}
\hline
Id. & Area & RA & Dec & FUV mag & g-band mag & FUV mag err & g-band mag err & FUV$-$g & Age \\
 & (pix) & (deg) & (deg) & (mag) & (mag) & (mag) & (mag) & (mag) & (Myr) \\
(1) & (2) & (3) & (4) & (5) & (6) & (7) & (8) & (9) & (10) \\\hline

17 & 63 & 49.6118 & -66.4567 & 19.45 & 17.69 & 0.10 & 0.00 & 1.75 & 268.04 \\
18 & 39 & 49.6182 & -66.4566 & 20.04 & 19.61 & 0.14 & 0.00 & 0.43 & 34.95 \\
108 & 46 & 49.6400 & -66.4630 & 19.37 & 19.32 & 0.09 & 0.00 & 0.04 & 20.09 \\
134 & 40 & 49.6494 & -66.4636 & 19.91 & 20.30 & 0.13 & 0.00 & -0.38 & 9.11 \\
135 & 22 & 49.6450 & -66.4635 & 20.27 & 20.49 & 0.14 & 0.00 & -0.22 & 11.57 \\
\hline
\end{tabular}
\tablecomments{This table shows a representative portion of the full catalog containing 1000 SFCs. The column descriptions are the same as in Table \ref{tab:catalog5457g}. }
\end{table}

\end{document}